# Title
Impact of Temperature and Relative Humidity on the Transmission of COVID-19: A Modeling Study in China and the United States

This paper was previously circulated under the title "**High Temperature and High Humidity Reduce the Transmission of COVID-19**"

Forthcoming  *BMJ Open*

## Authors
Jingyuan Wang[1,2], Ke Tang[3*], Kai Feng[1], Xin Lin[1], Weifeng Lv[1,4], Kun Chen[5,6] and Fei Wang[7]

## Affiliations
[1]School of Computer Science and Engineering, Beihang University, China.
[2]Beijing Advanced Innovation Center for Big Data and Brain Computing, Beihang University, China.
[3]School of Social Sciences, Tsinghua University, China.
[4]State Key Laboratory of Software Development Environment, Beihang University, China.
[5]Department of Statistics, University of Connecticut, U.S.
[6]Center for Population Health, University of Connecticut Health Center, U.S.
[7]Department of Population Health Sciences, Weill Cornell Medical College. Cornell University, U.S.

[*]Corresponding author: Ke Tang, School of Social Sciences, Tsinghua University, Beijing, China. Email: ketang@tsinghua.edu.cn

## ABSTRACT
**Objectives** We aim to assess the impact of temperature and relative humidity on the transmission of COVID-19 across communities after accounting for community-level factors such as demographics, socioeconomic status, and human mobility status.
**Design** A retrospective cross-sectional regression analysis via the Fama-MacBeth procedure is adopted.
**Setting** We use the data for COVID-19 daily symptom-onset cases for 100 Chinese cities and COVID-19 daily confirmed cases for 1,005 U.S. counties.
**Participants** A total of 69,498 cases in China and 740,843 cases in the U.S. are used for calculating the effective reproductive numbers.
**Primary outcome measures** Regression analysis of the impact of temperature and relative humidity on the effective reproductive number ($R$ value).
**Results** Statistically significant negative correlations are found between temperature/relative humidity and the effective reproductive number ($R$ value) in both China and the U.S.
**Conclusions** Higher temperature and higher relative humidity potentially suppress the transmission of COVID-19. Specifically, an increase in temperature by 1 degree Celsius is associated with a reduction in the $R$ value of COVID-19 by 0.026 (95% CI [-0.0395,-0.0125]) in China and by 0.020 (95% CI [-0.0311, -0.0096]) in the U.S.; an increase in relative humidity by 1% is associated with a reduction in the $R$ value by 0.0076 (95% CI [-0.0108,-0.0045]) in China and by 0.0080 (95% CI [-0.0150,-0.0010]) in the U.S. Therefore, the potential impact of temperature/relative humidity on the effective reproductive number alone is not strong enough to stop the pandemic.

**Strengths and limitations of this study**
1. Cross-sectional observations from 100 Chinese cities and 1,005 U.S. counties cover a wide spectrum of meteorological conditions.
2. Demographics, socioeconomic status, geographical, healthcare, and human mobility factors are all included in the regression analysis.
3. The Fama-MacBeth regression framework allows the identification of associations between temperature/relative humidity and COVID-19 transmissibility for nonstationary short-duration data.
4. The exact mechanism of the negative association between *R* and temperature/relative humidity has not been investigated in this study.
5. The temperature and relative humidity data collected from China and the U.S. do not contain extreme conditions.

**MAIN TEXT**

**Introduction**
The coronavirus disease 2019 (COVID-19) pandemic, caused by severe acute respiratory syndrome coronavirus 2 (SARS-CoV-2), has infected more than 70 million people with 1,595,187 deaths across 220 countries and territories as of December 13, 2020 [1], since its first reported case in Wuhan, China in December 2019 [2,3]. COVID-19 has had disastrous impacts on global public health, the environment, and socioeconomic status [4–7]. Understanding the factors that affect the transmission of SARS-CoV-2 is crucial for predicting the transmission dynamics of the virus and making appropriate intervention policies. Numerous recent studies have analyzed the effects of anthropogenic factors on COVID-19 transmission, such as travel restrictions [8–10], nonpharmacological interventions [11], population flow [12], anti-contagion policies [13], and contact patterns [14].

Meteorological factors, such as temperature and humidity, have previously been suggested to be associated with the transmissibility of certain infectious diseases. For example, prior studies have shown that the transmission of influenza is seasonal and is affected by humidity [15,16], and that wintertime climate and host behavior can facilitate the transmission of influenza [17–19]. Studies have also shown that the transmission of other human coronaviruses that cause mild respiratory symptoms, such as OC43 (HCoV-OC43) and HCoV-HKU1, is seasonal [20,21]. The seasonality of these related viruses has been leveraged in an indirect long-term simulation of the transmission of SARS-CoV-2 [22,23], and other studies have demonstrated a correlation between meteorological factors and pandemic spreading [24]. In addition, temperature and humidity have been shown to be important natural factors affecting pulmonary diseases [25], which are prevalent in COVID-19 patients.

However, there is no consensus on the impact of meteorological factors on COVID-19 transmissibility. For example, the study by Merow *et al.* shows that ultraviolet light is associated with a decreasing trend in COVID-19 case growth rates [26]. In contrast, other studies claim no association between COVID-19 transmissibility and temperature and ultraviolet light [27] or a positive association between temperature and daily confirmed cases [28,29]. Since the COVID-19 outbreak has lasted for less than a year, we do not have multiyear time-series data to estimate a stable serial cointegration between meteorological factors and certain indicators of COVID-19 transmissibility. As large-scale social intervention unfolded shortly after the outbreak in both countries, the periods without nonpharmaceutical intervention were quite short. Thus, estimation of the influences of meteorological factors on COVID-19 transmissibility is challenging.

The goal of this paper is to accurately quantify such influences, where the meteorological factors include temperature and humidity, and the COVID-19 transmissibility is measured by the effective reproductive number (*R* values). Our analysis is based on COVID-19 data from both China and the U.S. With several months of observations, the *R* values typically will have a trend, as will temperature and humidity. In this paper, we consider a strategy of "trading-space-for-time" by using Fama-MacBeth regression with Newey-West adjustment for standard errors, which is widely used in finance [30–32]. Specifically, we first estimate the cross-sectional association between temperature/relative humidity and *R* values across 100 cities in China from January 19 to February 15 (nationwide lockdown started from January 24) and 1,005 counties in the U.S. from March 15 to April 25 (nationwide lockdown started from April 7) and then adjust for the time-series autocorrelation of these estimates. Demographics, socioeconomic status, geographical, healthcare, and human mobility status factors are also included in our modeling process as control variables. Our framework enables analysis during the early stage of an infectious disease outbreak and thus has considerable potential for informing policymakers to consider social interventions in a timely fashion.

## Materials and Methods

**Data.**
Records of 69,498 COVID-19 patients with symptom-onset days up to February 10, 2020 from 325 cities are extracted from the Chinese National Notifiable Disease Reporting System. Each patient's records include the area code of his/her current residence, the area code of the reporting institution, the date of symptom onset and the date of confirmation. With such symptom-onset data, we are able to estimate the precise *R* values for different Chinese cities. For U.S. data, daily confirmed cases for 1,005 counties with a more than 20,000 population size are collected from the COVID-19 database of the Johns Hopkins University Center for Systems Science and Engineering (which is publicly available at https://github.com/CSSEGISandData/COVID-19/). We extract the data from March 15 to April 25 for the 1,005 counties, which results in a total of 740,843 confirmed cases. Due to the unavailability of onset date information in the U.S. data, we estimate *R* values from the daily confirmed cases for U.S. counties, which may be less precise than the estimation for the Chinese cities.

We also collect 4,711 cases from Chinese epidemiological surveys published online by the Centers for Disease Control and Prevention of 11 provinces and municipalities, including Beijing, Shanghai, Jilin, Sichuan, Hebei, Henan, Hunan, Guizhou, Chongqing, Hainan and Tianjin. By analyzing the records of each patient's contact history, we match close contacts and select 105 pairs of clear virus carriers and infections, which are used to estimate the serial intervals of COVID-19.

Temperature and relative humidity data are obtained from 699 meteorological stations in China from http://data.cma.cn/. Other factors, including population density, GDP per capita, the fraction of the population aged 65 and above, and the number of doctors for each city in 2018, are obtained from https://data.cnki.net. The indices indicating the number of migrants from Wuhan to other cities over the period of January 7 to February 10 and the Baidu Mobility Index are obtained from https://qianxi.baidu.com/. Panel A of Table S1 in the supplementary materials provides the summary statistics of the variables for analyzing the data from China with their pairwise correlations shown in Table S2.

For the U.S., temperature and relative humidity data are collected from the National Oceanic and Atmospheric Administration (https://www.ncdc.noaa.gov/). Population data and the fraction of residents over 65 years of age for each county are obtained from the American Community Survey (https://www.census.gov/). GDP and personal income in 2018 for each county are obtained from https://www.bea.gov/. Data describing mobility changes, including the fraction of maximum moving distance over normal time and home-stay minutes for each county, are obtained from https://github.com/descarteslabs/DL-COVID-19 and https://www.safegraph.com/. The Gini index, the fraction of the population below the poverty level, the fraction of residents who are not in the labor force (under 16 years old), the fraction of households with a total income greater than $200,000, and the fraction of the population with food stamp/SNAP benefits are obtained from the American Community Survey. The number of ICU beds for each county is obtained from https://www.kaggle.com/jaimeblasco/icu-beds-by-county-in-the-us/data. Panel B of Table S1 in the supplementary materials provides the summary statistics of the variables for analyzing the U.S. data with their pairwise correlations shown in Table S3.

**Patient and public involvement**
In this study, in order to protect the patients' privacy, no identifiable protected health information is extracted from the Chinese National Notifiable Disease Reporting System. The Chinese epidemiological surveys data has personal information removed before publication. Patient and/or public are not involved in the design, or conduct, or reporting, or dissemination plans of this research.

**Construction of Effective Reproductive Numbers**.
We use the effective reproductive number, or the *R* value, to quantify the transmission of COVID-19 in different cities and counties. The calculation of the *R* value consists of two steps. First, we estimate the serial interval, which is the time between successive cases in a transmission chain of COVID-19 using 105 pairs of virus carriers and infections. We fit these 105 samples of serial intervals with a Weibull distribution using maximum likelihood estimation (MLE) (implemented with the Python package 'Scipy' and R package 'MASS' (Python version 3.7.4, 'Scipy' version 1.3.1 and R version 3.6.2, 'MASS' version 7.3_51.4)), as shown in Figure S1. The results of the two implementations are consistent with each other. The mean and standard deviation of the serial intervals are 7.4 and 5.2 days, respectively.

Note that cities with a small number of confirmed cases typically have a highly wiggy *R* value curve due to inaccurate *R* value estimation. Therefore, we select cities with more than 40 cases in China, 100 in total. We then calculate the *R* value for each of the 100 Chinese cities from the date of the first-case to February 10 through a time-dependent method based on MLE (Supplementary Materials pages 4-5) [33]. For estimation of *R* values in U.S. counties, the settings of serial intervals are set to the same as China, *i.e.,* with a 7.4 day mean and 5.2 day standard deviation. We use the same methods of estimating the *R* values of all 1,005 U.S. counties from the date when the first confirmed case occurred in the county to April 25, 2020.

**Study Period**.
We aim to study the influences of various factors on the *R* value under the outdoor environment, because if people stay at home for most of their time under the restrictions of the isolation policy, weather conditions are unlikely to influence virus transmission. We thus perform separate analyses before and after the large-scale stay-at-home quarantine policies for both China (January 24) and the U.S. (April 7). The first-level response to major public health emergencies in many major Chinese cities and provinces, including Beijing and Shanghai, was announced on January 24. Moreover, the numbers of cases in most cities before January 18 are too small to accurately estimate the *R* value. Therefore, we take the daily *R* values from January 19 to January 23 for each city as the before-lockdown period. Although Wuhan City imposed a travel restriction at 10 a.m. on January 23, a large number of people still left Wuhan before 10 a.m. on that day, so our sample still includes January 23 for Wuhan. We take January 24 to February 10 as the period after lockdown for China. As reported by The New York Times, most states announced state-wide stay-at-home orders from April 7 for the U.S. [34]. Moreover, the number of cases in most counties before March 15 is too small to accurately estimate the *R* value, so we take March 15 to April 6 for each county as the before-lockdown period and April 7 to April 25 as the after-lockdown period.

**Statistical Analysis**.
We use six-day average temperature and relative humidity values up to and including the day when the $R$ value is measured. Our strategy is inspired by the five-day incubation period estimated from Johns Hopkins University [35] plus a one-day onset. In the data of this work, the series of the 6-day average temperature and relative humidity and the daily $R$ values are mostly nonstationary. We find a declining trend of $R$ values for nearly all Chinese cities and the U.S. counties during our study periods, which could be due to the nature of the disease and people's raised awareness and increased self-protection measures even before the lockdown. Table S4 Panel A and Panel B in the supplementary materials show the panel Handri LM unit root test [36] results for the China and U.S. data. In this case, direct time-series regression cannot be applied due to the so-called spurious regression [37] problem, which states the fact that a regression may provide misleading statistical evidence of a linear relationship between nonstationary time-series variables. We thus adopt the Fama-MacBeth methodology [38] with Newey-West adjustment, which consists of a series of cross-sectional regressions and has been proven effective in various disciplines, including finance and economics. The details are described as follows.

**Fama-MacBeth Regression with the Newey–West Adjustment**.
Fama-MacBeth regression is a two-step procedure (Supplementary Materials p2-3). In the first step, it runs a cross-sectional regression at each point in time; the second step estimates the coefficient as the average of the cross-sectional regression estimates. Since these estimates might have autocorrelations, we adjust the error of the average with a Newey-West approach. Mathematically, our method proceeds as follows.

Step 1: Let $T$ be the length of the time period and $M$ be the number of control variables. For each timestamp $t$, we run a cross-sectional regression:
$$R_{i,t} = c_t + \beta_{temp,t} * temp_{i,t} + \beta_{humi,t} * humi_{i,t} + \sum_{j=1}^{M} \beta_{control_j,t} * control_{j,i,t} + \epsilon_{i,t}$$
Step 2: Estimate the average of the regression coefficient estimates obtained from the first step:
$$\hat{\beta}_k = \frac{1}{T}\sum_{t=1}^{T} \beta_{k,t}$$

We use the Newey-West approach [39] to adjust for the time-series autocorrelation and heteroscedasticity in calculating the standard errors in the second step. Specifically, the Newey-West estimators can be expressed as
$$S = \frac{1}{T}(\sum_{t=1}^{T} e_t^2 + \sum_{l=1}^{L}\sum_{t=l+1}^{T} w_l e_t e_{t-l}),$$
where $w_l = 1 - \frac{l}{1+L}$, where $e$ represents residuals and $L$ is the lag (Supplementary Materials pages 2-3).

The Fama-MacBeth regression with Newey-West adjustment has two advantages: 1) It avoids the spurious regression problem for nonstationary series, as the first-step estimates, $\{\beta_{k,t}\}$, have much milder autocorrelations than the autocorrelations (time trends) within the observations. Such autocorrelations can be adjusted by the Newey-West procedure. 2) Only cross-sectional coefficient estimates in the first step are used to estimate the coefficients, but not their standard errors; hence, any heteroskedasticity and residual-dependent issues in the first step will not influence the final results, because the heteroskedasticity and residual dependency (including the one caused by spatial correlation) does not alter the unbiasedness of the coefficient in the ordinary least squares (OLS) estimation. Table S5 shows the detailed coefficients of temperature and relative humidity in the first step of the Fama-MacBeth regression.

Note that the Fama-MacBeth regression with Newey-West adjustment is commonly used in estimating parameters for finance and economic models that are valid in the presence of cross-sectional correlation and time-series autocorrelation [30–32]. To the best of our knowledge, our study is a novel application of this method in emergent public health and epidemiological problems.

In our implementation, on each day of the study period, we perform a cross-sectional regression of the daily $R$ values of various cities or counties based on their 6-day average temperature and relative humidity values, as well as several categories of control variables, including the following:
(1) *Demographics.* The population density and the fraction of people aged 65 and older for both China and the U.S.
(2) *Socioeconomic statuses.* The GDP per capita for Chinese cities. For the U.S. counties, the Gini index and the first PCA factor derived from several factors including GDP per capita, personal income, the fraction of the population below the poverty level, the fraction of the population not in the labor force (16 years or over), the fraction of the population with a total household income more than $200,000, and the fraction of the population with food stamp/SNAP benefits.
(3) *Geographical variables.* Latitudes and longitudes.
(4) *Healthcare.* The number of doctors in Chinese cities and the number of ICU beds per capita for U.S. counties.
(5) *Human mobility status.* For Chinese cities, the number of people that migrated from Wuhan in the 14 days prior to the $R$ measurement and the drop rate of the Baidu Mobility Index compared to the same day in the first week of Jan 2020. For U.S. counties, the fraction of maximum moving distance over the median of normal time (weekdays from Feb 17 to March 7), and home-stay minutes are used as mobility proxies. All human mobility controls are averaged over a 6-day period in the regression.

All analyses are conducted in Stata version 16.0.

**Results**

COVID-19 has spread widely in both China and the U.S. The transmissibility and meteorological conditions in the cities/counties of these two countries vary greatly (see Figures 1 and 2). We analyze the relationship between COVID-19 transmissibility and temperature/relative humidity, controlling for various demographics, socioeconomic statuses, geographical, healthcare, and human mobility status factors and correcting for cross-sectional correlations. Overall, we find robust negative correlations between COVID-19 transmissibility before the large-scale public health interventions (lockdown) in China and the U.S. and temperature and relative humidity. Moreover, temperature has a consistent influence on the effective reproductive number, $R$ values, for both Chinese cities and U.S. counties; relative humidity also has consistent effects across the two countries. Both of them continue to have a negative influence even after the public health intervention, but with smaller magnitudes since an increasing number of people stay at home and hence are exposed less to the outdoor weather. More details are presented below.

**Temperature, Relative Humidity, and Effective Reproductive Numbers**

For both China and the U.S., we conduct a series of cross-sectional regressions (the Fama-MacBeth approach [38]) of the daily effective reproductive numbers ($R$ values), which measure COVID-19 transmissibility, on the six-day average temperature and relative humidity up to and including the day when the $R$ value is measured, considering the transmission during presymptomatic periods [35] and other control factors for the before-lockdown period, the after-lockdown period, and the overall period. Figure 1 shows the average $R$ values from January 19 to 23 (before lockdown) for different Chinese cities geographically, and Figure 2 shows the average $R$ values from March 15 to April 6 (before the majority of states declared a stay-at-home order) for different U.S. counties.

Overall, the results for Chinese cities (Table 1) demonstrate that the six-day average temperature and relative humidity have a significant relationship with $R$ values, with p-values smaller than or approximately 0.01 for all three specified time periods. The analysis of U.S. counties (Table 2) shows that six-day average temperature and relative humidity have statistically significant correlations with *R* values, with p-values lower than 0.05 before April 7, the time when most states declared state-wide stay-at-home orders [34].

The influences of the temperature and relative humidity on the *R* values are quite similar before the lockdown in China and the U.S.: a one-degree Celsius increase in temperature is associated with an approximately 0.023 decrease (-0.026 (95% CI [-0.0395,-0.0125]) in China and -0.020 (95% CI [-0.0311, -0.0096]) in the U.S.) in the *R* value, and a one percent relative humidity rise is associated with an approximately 0.0078 decrease (-0.0076 (95% CI [-0.0108,-0.0045]) in China and -0.0080 (95% CI [-0.0150,-0.0010]) in the U.S.) in the *R* value. After lockdown, the temperature and relative humidity also present negative relationships with the *R* values for both countries. For China, it is statistically significant (with p-values lower than 0.05), and a one-degree Celsius increase in temperature and a one percent increase in relative humidity are associated with a 0.0209 decrease (95% CI [-0.0378, -0.0041]) and a 0.0054 decrease (95% CI [-0.0104, -0.0004]) in the *R* value, respectively. For the U.S., the estimated effects of temperature and relative humidity on the *R* values are still negative but no longer statistically significant (with p-values of 0.141 and 0.073, respectively). The lesser influence of weather conditions is very likely caused by the stay-at-home policy during lockdown periods, when people are less exposed to the outdoor weather. Therefore, we rely more on the estimates of the weather-transmissibility relationship before the lockdowns in both countries.

**Control Variables.**
Several control variables also have significant influences on COVID-19 transmissibility. In China, before the lockdowns, in cities with higher levels of population density, the virus spreads faster than in less crowded cities due to more possible contacts among people. A one thousand people per square kilometer increase in population density is associated with a 0.1188 increase (95% CI [0.0573, 0.1803]) in the *R* value before lockdown. Cities in China with more doctors have a smaller transmission intensity since the infections are treated in hospitals and hence are unable to be transmitted to others. In particular, one thousand more doctors are associated with a 0.0058 decrease (95% CI [-0.0090, -0.0025]) in the *R* value during the overall time period; the influence of doctor number is greater before lockdown with a coefficient of 0.0109 (95% CI [-0.0163, -0.0056])). Similarly, more developed cities (with higher GDP per capita) normally have better medical conditions; hence, patients are more likely to be cared for and thus unlikely to be transmitting the infection to others. A ten thousand Chinese Yuan GDP per capita increase is associated with a decrease in the *R* value by 0.0145 (95% CI [-0.0249, -0.0040]) before the lockdown. In the U.S., there is a strong relationship between the *R* value and the number of ICU beds per capita after lockdown, with a p-value of 0.001; every unit increase in ICU bed per 10,000 population is associated with a 0.0110 decrease (95% CI [-0.0171, -0.0049]) in the *R* value. Moreover, counties with more people over 65 years old have lower *R* values, but the magnitude is small, *i.e.,* a one percent increase in the fraction of individuals aged over 65 is associated with a 0.0092 decrease (95% CI [-0.0135, -0.00498]) in the *R* value in the overall time period.

**Absolute Humidity.**
Absolute humidity, the mass of water vapor per cubic meter of air, relates to both temperature and relative humidity. A previous work shows that absolute humidity is a good solo variable explaining the seasonality of influenza [40]. The results shown in Table 3 are only partly consistent with this notion [40]. In particular, for the U.S. counties, relative humidity and absolute humidity are almost equivalent in explaining the variation in the *R* value (12.57% vs. 12.55%), while absolute humidity does achieve a higher significance level (p-value less than 0.00001) than relative humidity (p-value of 0.019) before lockdown. However, the coefficient of absolute humidity is not statistically significant for Chinese cities (p-value of 0.312).

**Lockdown and Mobility**.

Intensive health emergency and lockdown policies have taken place since the outbreak of COVID-19 in both the U.S. and China. In the regression analysis, we use cross-sectional centralized (with sample mean extracted) explanatory variables, and thus, the intercepts in the regression models estimate the average $R$ value of different time periods. In China, the health emergency policies on January 24, 2020 lowered the average $R$ value from 2.1174 (95% CI [1.5699, 2.6649]) to 0.8084 (95% CI [0.5334, 1.0833]), which corresponds to a more than 60% drop. In the U.S., the regression results of the data as of April 25 show that although the $R$ value has not decreased to less than 1, the lockdown policies have reduced the average $R$ value by nearly half, from 2.1970 (95% CI [1.6631, 2.7309]) to 1.1837 (95% CI [1.1687, 1.1985]).

We use the Baidu Mobility Index (BMI) drop as a proxy for intracity mobility change (compared to the normal time) in China. The regression results show that before the lockdown, a 1% decrease in BMI drop is associated with a decrease in the $R$ value by 0.004093 (95% CI [-0.00683, -0.001356]). After the lockdown, the BMI drop does not significantly affect the $R$ value. A possible reason is that the BMI variations across cities are quite small (all at quite low levels) after the lockdown, as the paces of interventions in different Chinese cities are quite similar. Overall, the negative relationship before lockdown may also imply that the rapid response to infectious disease risks is crucial. For the U.S., we use the M50 index, the fraction of daily median of maximum moving distance over that in the normal time (workdays between February 17 and March 7), as the proxy of mobility. It has a positive relationship with the $R$ value both overall and after-lockdown time period, with p-values lower than 0.01, which demonstrates that counties with more social movements would have higher $R$ values than others.

**Robustness Checks**.

We check the robustness of the influences of temperature/humidity on $R$ values over four conditions:

(1) **Wuhan city.** Among these 100 cities in China, Wuhan is a special case with the earliest outbreak of COVID-19. There was an increase of more than 13,000 cases on a single day (February 12, 2020) due to the unification of testing standards with other regions of China [41]. Therefore, as a robustness check, we remove Wuhan city from our sample and redo the regression analysis.
(2) **Different measurements of serial intervals.** We also use serial intervals in a previous work (mean 7.5 days, std 3.4 days based on 10 cases) [3] with a Weibull distribution to estimate the $R$ values of various cities/counties for robustness checks.
(3) **Social distancing dummy variables for the U.S. counties.** States in the U.S. announced stay-at-home orders at different times. We add a dummy variable that is set to one if the stay-at-home order is imposed and zero otherwise.
(4) **Spatial random effect.** We also introduce a spatial model into the first step of the Fama-MacBeth regression to account for spatial correlation and redo the analysis.

The results of the abovementioned four robustness checks are shown in Table S6 to S11. All of them show that temperature and relative humidity have a strong influence on $R$ values with strong statistical significance, which is consistent with the reported results in Tables 1 and 2.

**Discussion**

We identify robust negative correlations between temperature/relative humidity and the COVID-19 transmissibility using samples of the daily transmission of COVID-19, temperature and relative humidity for 100 Chinese cities and 1,005 U.S. counties. Although we use different datasets (symptom-onset data for Chinese cities and confirmed case data for the U.S. counties) for different countries, we obtain consistent estimates. This result also aligns with the evidence that high temperature and high humidity can reduce the transmission of influenza [40], which can be explained by several potential reasons. The influenza virus is more stable in cold environments, and respiratory droplets, as containers of viruses, remain airborne longer in dry air [42]. Cold and dry weather can also weaken host immunity and make the hosts more susceptible to the virus [43]. Our result is also consistent with the evidence that high temperature and high relative humidity reduce the viability of SARS coronavirus [44]. High transmission in cold temperatures may also be explained by behavioral differences; for instance, people may spend more time indoors and have a greater chance of interacting with others. Further studies should be performed to disentangle these multiple explanations and change the association relationship in our study to a causal effect.

Our study has several strengths. First, we use data from vast geographical scopes in both China and the U.S. that contain a variety of meteorological conditions. Second, we employ all kinds of control variables such as demographics, socioeconomic status, geographical, healthcare and human mobility status factors as control variables to capture the effect of regional disparity. Third, we use the Fama-MacBeth regression framework to estimate associations between temperature/relative humidity and COVID-19 transmissibility when our data are nonstationary and in a short duration. Compared to the study by Merow *et al.,* which investigates the influence of meteorological conditions on COVID-19 infections with only population density and the proportion of individuals aged over 65 years considered as control variables [26], our study incorporates more categories of variables to explain the heterogeneity among different regions. Although a study by Yao *et al.* has announced no association between COVID-19 transmission and temperature, they use a 2-month averaged temperature for analysis, and the temperature trends are not considered [27]. A study by Xie *et al.* reports positive relationships between temperature and COVID-19 cases [29]. However, the demographic factors for cities are not incorporated as controls, and the effectiveness of nonstationary time series problem for the panel regression methods they use is not explicitly discussed.

We do acknowledge several limitations. Our findings cannot verify the detailed mechanisms between temperature/relative humidity and COVID-19 transmissibility. Our study is a statistical analysis but not an experiment. These findings should be considered with caution when used for prediction. The $R^2$ of our regression is approximately 30% in China and 12% in the U.S., which means that approximately 70% to 88% of cross-city *R* value fluctuations cannot be explained by temperature and relative humidity (and controls). Moreover, the temperatures and relative humidity in our Chinese samples range from -21°C to 20°C and from 49% to 100%, respectively, and in the U.S., the temperature and humidity range from -10°C to 29°C and from 16% to 99%, respectively; thus, it is still unknown whether these negative relationships still hold in extremely hot and cold areas. The slight differences between the estimates on the Chinese cities and the U.S. counties might come from the different ranges of temperature and relative humidity.

Outwardly, our study suggests that the summer and rainy seasons can potentially reduce the transmissibility of COVID-19, but it is unlikely that the COVID-19 pandemic will "automatically" diminish in summer. Cold and dry seasons can potentially break the fragile transmission balance and the weaken downward trends in some areas of the Northern Hemisphere.

Therefore, public health intervention is still necessary to block the transmission of COVID-19 even in the summer. In particular, as shown in this paper, lockdowns, constraints on human mobility, and increases in hospital beds, can potentially reduce the transmissibility of COVID-19. Given the relationship between temperature/relative humidity and COVID-19 transmissibility, policymakers can adjust their intervention policy according to the different temperature/relative humidity conditions. When new infectious diseases emerge, our framework can also provide policymakers with fast support, although this is not expected.

**Contributorship statement** J.W. initiated this project. J.W., W.L. and F.W. planned and oversaw the project. K.T. and K.C. contributed econometrics methods. K.F and X.L. prepared the datatsets and conducted analysis. K.T, W.F and J.W. wrote the manuscript with input from all authors.

**Competing interests** The authors declare no competing interests.

**Funding** This study was granted the State Key Research and Development Program of China (2019YFB2102100), the National Natural Science Foundation of China (92046010, 71531001, 61572059, 61421003), and BRICS STI Framework Programme: Response to COVID-19 global pandemic(MFQuantiC).

**Data sharing statement** Data are available on reasonable request. Temperature, humidity, R values and all control variables except home-stay minutes used in this study are available on request from JW (jywang@buaa.edu. cn). Home-stay minute data provided by Safegraph (https://www.safegraph.com/) cannot be disclosed since this would compromise the agreement with the data provider; nevertheless, these data can be obtained by applying for permission from the provider

# Figures and Tables

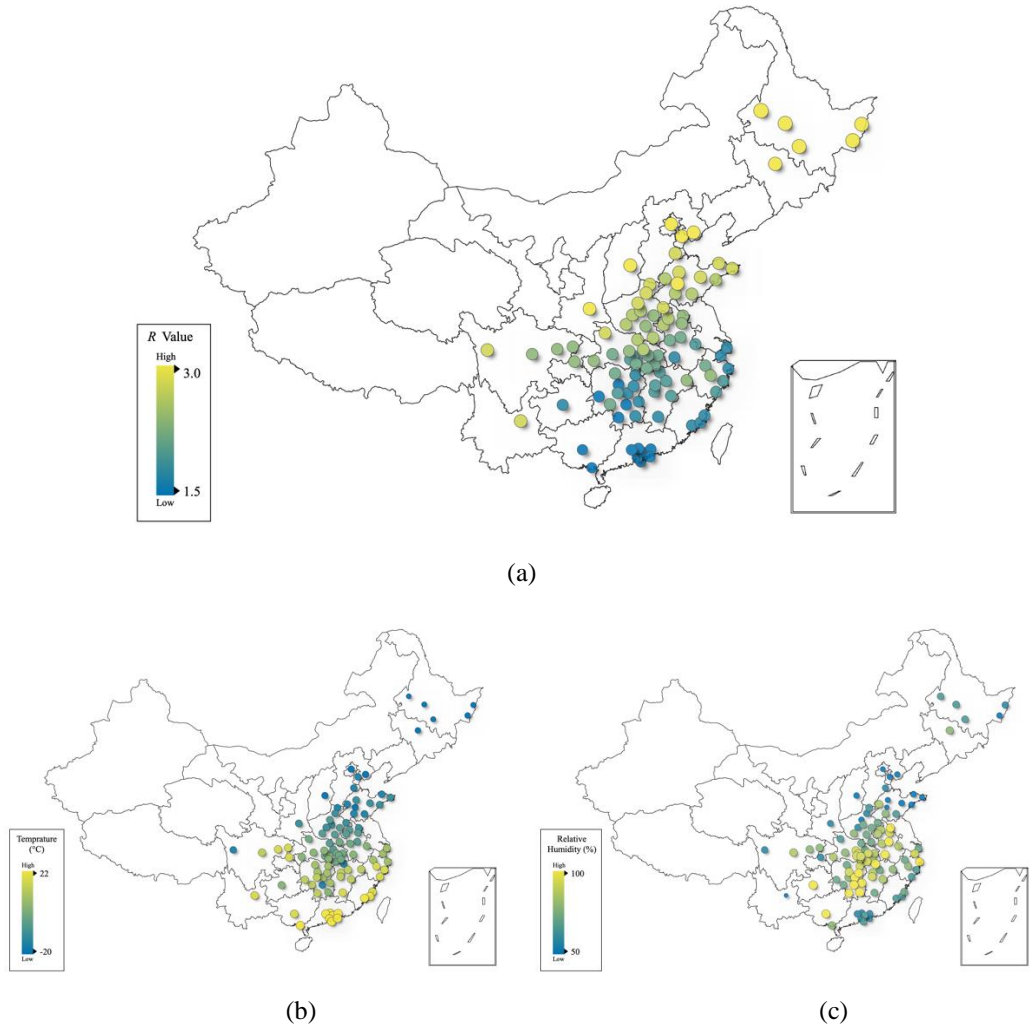

**Figure 1: A city-level visualization of COVID-19 transmission (a), temperature (b) and relative humidity (c).**
Average $R$ values from January 19 to 23, 2020 for 100 Chinese cities are used in subplot (a). The average temperature and relative humidity for the same period are plotted in (b) and (c).

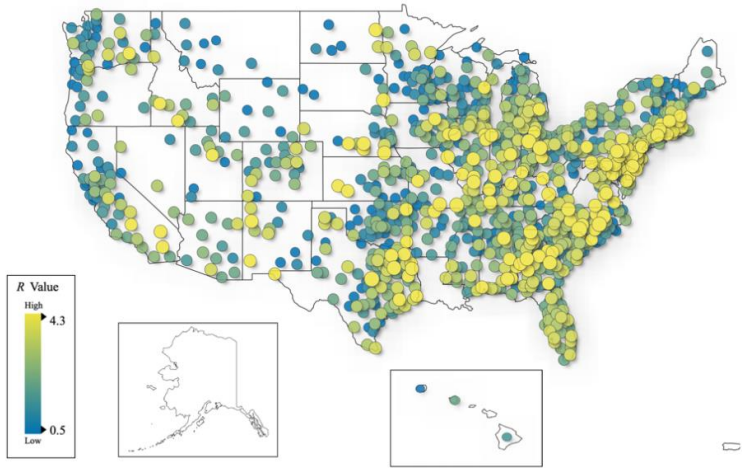

(a)

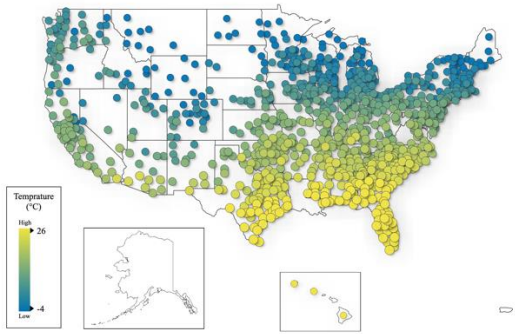

(b)

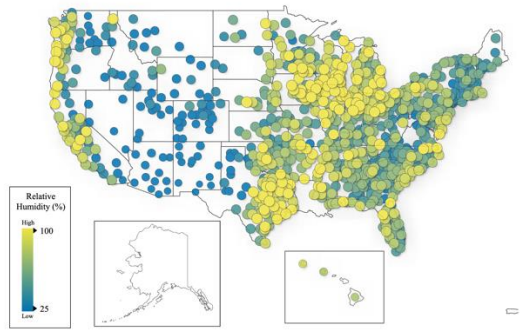

(c)

**Figure 2: A county-level visualization of COVID-19 transmission (a), temperature (b) and relative humidity (c) in the U.S.**
Average *R* values from March 15 to April 6, 2020 for 1,005 U.S. counties are used in subplot (a). The average temperature and relative humidity for the same period are plotted in (b) and (c).

## Table 1: Fama-MacBeth Regression for Chinese Cities

Daily $R$ values from January 19 to February 10 and averaged temperature and relative humidity over 6 days up to and including the day when $R$ value is measured, are used in the regression for 100 Chinese cities with more than 40 cases. The regression is estimated by the Fama-MacBeth approach.

|  | Overall | Before Lockdown (Jan 24) | After Lockdown (Jan 24) |
|---|---|---|---|
| $R^2$ | 0.3013 | 0.1895 | 0.3323 |
| **Temperature** | | | |
| coef | -0.0220 | -0.0260 | -0.0209 |
| 95%CI | [-0.0356,-0.0085] | [-0.0395,-0.0125] | [-0.0378,-0.0041] |
| std.err | 0.0065 | 0.0049 | 0.0080 |
| t-stat | -3.38 | -5.35 | -2.62 |
| p-value | 0.003 | 0.006 | 0.018 |
| **Relative Humidity** | | | |
| coef | -0.0059 | -0.0076 | -0.0054 |
| 95%CI | [-0.0098,-0.0019] | [-0.0108,-0.0045] | [-0.0104,-0.0004] |
| std.err | 0.0019 | 0.0011 | 0.0024 |
| t-stat | -3.08 | -6.70 | -2.29 |
| p-value | 0.005 | 0.003 | 0.035 |
| **Population Density** | | | |
| coef | 0.0259 | 0.1188 | 0.0001 |
| 95%CI | [-0.0292,0.0810] | [0.0573,0.1803] | [-0.0359,0.0362] |
| std.err | 0.0266 | 0.0222 | 0.0171 |
| t-stat | 0.98 | 5.36 | 0.01 |
| p-value | 0.340 | 0.006 | 0.993 |
| **Percentage over 65** | | | |
| coef | 0.1255 | 0.3230 | 0.0707 |
| 95%CI | [-1.7524,2.0034] | [-1.1797,1.8256] | [-2.3231,2.4644] |
| std.err | 0.9055 | 0.5412 | 1.1346 |
| t-stat | 0.14 | 0.60 | 0.06 |
| p-value | 0.891 | 0.583 | 0.951 |
| **GDP per capita** | | | |
| coef | 0.0045 | -0.0145 | 0.0098 |
| 95%CI | [-0.0157,0.0248] | [-0.0249,-0.0040] | [-0.0105,0.0301] |
| std.err | 0.0098 | 0.0038 | 0.0096 |
| t-stat | 0.46 | -3.85 | 1.02 |
| p-value | 0.647 | 0.018 | 0.322 |
| **No. of doctors** | | | |
| coef | -0.0058 | -0.0109 | -0.0043 |
| 95%CI | [-0.0090,-0.0025] | [-0.0163,-0.0056] | [-0.0064,-0.0022] |
| std.err | 0.0015 | 0.0019 | 0.0010 |
| t-stat | -3.71 | -5.69 | -4.41 |

|  | Overall | Before Lockdown (Jan 24) | After Lockdown (Jan 24) |
|---|---|---|---|
| p-value | 0.001 | 0.005 | 0.0004 |
| **Drop of BMI** | | | |
| coef | 0.3051 | -0.4093 | 0.5036 |
| 95%CI | [-0.3352,0.9454] | [-0.6830,-0.1356] | [-0.1133,1.1205] |
| std.err | 0.3087 | 0.0986 | 0.2924 |
| t-stat | 0.99 | -4.15 | 1.72 |
| p-value | 0.334 | 0.014 | 0.103 |
| **Inflow population from Wuhan** | | | |
| coef | -0.0052 | -0.0006 | -0.0065 |
| 95%CI | [-0.0106,0.0002] | [-0.0010,-0.0001] | [-0.0127,-0.0003] |
| std.err | 0.0026 | 0.0002 | 0.0029 |
| t-stat | -2.00 | -3.58 | -2.21 |
| p-value | 0.058 | 0.023 | 0.041 |
| **Latitude** | | | |
| coef | 0.0046 | 0.0096 | 0.0032 |
| 95%CI | [-0.0145,0.0236] | [-0.0133,0.0325] | [-0.0211,0.0274] |
| std.err | 0.0092 | 0.0083 | 0.0115 |
| t-stat | 0.50 | 1.16 | 0.28 |
| p-value | 0.625 | 0.311 | 0.786 |
| **Longitude** | | | |
| coef | -0.011 | -0.0270 | -0.0065 |
| 95%CI | [-0.0199,-0.0021] | [-0.0528,-0.0013] | [-0.0137,0.0007] |
| std.err | 0.0043 | 0.0093 | 0.0034 |
| t-stat | -2.56 | -2.92 | -1.91 |
| p-value | 0.018 | 0.043 | 0.074 |
| **const** | | | |
| coef | 1.0929 | 2.1174 | 0.8084 |
| 95%CI | [0.5078,1.6781] | [1.5699,2.6649] | [0.5334,1.0833] |
| std.err | 0.2821 | 0.1972 | 0.1303 |
| t-stat | 3.87 | 10.74 | 6.20 |
| p-value | 0.001 | 0.0004 | 0 |

## Table 2: Fama-MacBeth Regression for the U.S. Counties

Daily *R* values from March 15 to April 25 and temperature and relative humidity over 6 days up to and including the day when *R* value is measured, are used in the regression for 1,005 U.S. counties with more than 20,000 population. The regression is estimated by the Fama-MacBeth approach.

|  | Overall | Before Lockdown (April 7) | After Lockdown (April 7) |
|---|---|---|---|
| R2 | 0.1155 | 0.1344 | 0.0925 |
| **Temperature** | | | |
| coef | -0.0165 | -0.0204 | -0.0118 |
| 95%CI | [-0.0257,-0.0073] | [-0.0311,-0.0096] | [-0.0279,0.0043] |
| std.err | 0.0045 | 0.0052 | 0.0077 |
| t-stat | -3.62 | -3.93 | -1.54 |
| p-value | 0.001 | 0.001 | 0.141 |
| **Relative Humidity** | | | |
| coef | -0.0049 | -0.0080 | -0.0013 |
| 95%CI | [-.0.0103,0.0005] | [-0.0150,-0.0010] | [-0.0027,0.0001] |
| std.err | 0.0027 | 0.0034 | 0.0007 |
| t-stat | -1.84 | -2.36 | -1.90 |
| p-value | 0.073 | 0.028 | 0.073 |
| **Population Density** | | | |
| coef | 4.39E-6 | 7.00E-6 | 1.23E-6 |
| 95%CI | [-0.00001,0.00002] | [-0.00003,0.00004] | [9.84E-7,3.45E-6] |
| std.err | 8.44E-6 | 0.00002 | 1.05E-6 |
| t-stat | 0.52 | 0.44 | 1.17 |
| p-value | 0.606 | 0.666 | 0.258 |
| **Percentage over 65** | | | |
| coef | -0.9243 | -1.1084 | -0.7014 |
| 95%CI | [-1.3510,-0.4976] | [-1.8119,-0.4050] | [-1.0696,-0.3332] |
| std.err | 0.2113 | 0.3392 | 0.1752 |
| t-stat | -4.37 | -3.27 | -4.00 |
| p-value | 0.0001 | 0.004 | 0.001 |
| **Gini** | | | |
| coef | -1.8428 | -1.9255 | -1.7426 |
| 95%CI | [-3.5058,-0.1797] | [-4.4539,0.6028] | [-2.4697,-1.0154] |
| std.err | 0.8235 | 1.2191 | 0.3461 |
| t-stat | -2.24 | -1.58 | -5.03 |
| p-value | 0.031 | 0.129 | 0.0001 |
| **Socio-economic factor** | | | |
| coef | 0.0916 | 0.1406 | 0.0324 |
| 95%CI | [0.0338,0.1495] | [0.0886,0.1925] | [-0.0108,0.0756] |
| std.err | 0.0287 | 0.0250 | 0.0206 |
| t-stat | 3.20 | 5.61 | 1.58 |

| | Overall | Before Lockdown (April 7) | After Lockdown (April 7) |
|---|---|---|---|
| p-value | 0.003 | 0.00001 | 0.133 |
| **No. of ICU beds per capita** | | | |
| coef | -0.0097 | -0.0086 | -0.0110 |
| 95%CI | [-0.0233,0.0039] | [-0.0299,0.0126] | [-0.0171,-0.0049] |
| std.err | 0.0067 | 0.0102 | 0.0029 |
| t-stat | -1.44 | -0.84 | -3.81 |
| p-value | 0.156 | 0.408 | 0.001 |
| **Fraction of maximum moving distance over normal time** | | | |
| coef | 0.0038 | 0.0022 | 0.0057 |
| 95%CI | [0.0014,0.0062] | [-0.0008,0.0053] | [0.0048,0.0066] |
| std.err | 0.0012 | 0.0015 | 0.0004 |
| t-stat | 3.23 | 1.50 | 13.71 |
| p-value | 0.002 | 0.147 | 0 |
| **Home stay minutes** | | | |
| coef | 0.0003 | 0.0008 | -0.0002 |
| 95%CI | [-0.0002,0.0008] | [0.0004,0.0011] | [-0.0004, -0.00003] |
| std.err | 0.0002 | 0.0002 | 0.0001 |
| t-stat | 1.32 | 4.46 | -2.40 |
| p-value | 0.194 | 0.0002 | 0.027 |
| **Latitude** | | | |
| coef | -0.0174 | -0.0333 | 0.0018 |
| 95%CI | [-0.0357,0.0009] | [-0.0492,-0.0173] | [-0.0189,0.0224] |
| std.err | 0.0091 | 0.0077 | 0.0098 |
| t-stat | -1.92 | -4.33 | 0.18 |
| p-value | 0.061 | 0.0003 | 0.861 |
| **Longitude** | | | |
| coef | 0.0068 | 0.0102 | 0.0027 |
| 95%CI | [0.0031,0.0105] | [0.0082,0.0122] | [0.0004,0.0049] |
| std.err | 0.0018 | 0.0010 | 0.0011 |
| t-stat | 3.71 | 10.51 | 2.49 |
| p-value | 0.001 | 0 | 0.023 |
| **const** | | | |
| coef | 1.7386 | 2.1970 | 1.1837 |
| 95%CI | [1.1784,2.2988] | [1.6631,2.7309] | [1.1687,1.1985] |
| std.err | 0.2774 | 0.2574 | 0.0071 |
| t-stat | 6.27 | 8.53 | 166.63 |
| p-value | 0 | 0 | 0 |

# Table 3: Absolute Humidity

Table 3 shows the explanatory power of the absolute humidity in the pre-lockdown period for Chinese cities from January 19 to 23 (Panel A) and the U.S. counties from March 15 to April 6 (Panel B).

## Panel A: Regression for Chinese Cities

|  | Temperature | Relative Humidity | Absolute Humidity |
|---|---|---|---|
| **R2** | 0.1817 | 0.1783 | 0.1799 |
| **Temperature** | | | |
| coef | -0.0151 | | |
| 95%CI | [-0.0262, -0.0040] | | |
| std.err | 0.0040 | | |
| t-stat | -3.78 | | |
| p-value | 0.019 | | |
| **Relative Humidity** | | | |
| coef | | -0.0038 | |
| 95%CI | | [-0.0060, -0.0016] | |
| std.err | | 0.0008 | |
| t-stat | | -4.83 | |
| p-value | | 0.008 | |
| **Absolute Humidity** | | | |
| coef | | | -0.0159 |
| 95%CI | | | [-0.0545, 0.0227] |
| std.err | | | 0.0139 |
| t-stat | | | -1.15 |
| p-value | | | 0.316 |
| **Population Density** | | | |
| coef | 0.1222 | 0.1062 | 0.1190 |
| 95%CI | [0.0500, 0.1943] | [0.0441, 0.1684] | [0.0371, 0.2010] |
| std.err | 0.0260 | 0.0224 | 0.0295 |
| t-stat | 4.70 | 4.74 | 4.03 |
| p-value | 0.009 | 0.009 | 0.016 |
| **Percentage over 65** | | | |
| coef | -0.3769 | -0.5738 | -0.8898 |
| 95%CI | [-1.6135, 0.8597] | [-1.6715, 0.5239] | [-1.9335, 0.1538] |
| std.err | 0.4454 | 0.3954 | 0.3759 |
| t-stat | -0.85 | -1.45 | -2.37 |
| p-value | 0.445 | 0.220 | 0.077 |
| **GDP per capita** | | | |
| coef | -0.0174 | -0.0190 | -0.0205 |
| 95%CI | [-0.0303, -0.0046] | [-0.0328, -0.0052] | [-0.0340, -0.0069] |
| std.err | 0.0046 | 0.0050 | 0.0049 |
| t-stat | -3.76 | -3.81 | -4.20 |

|  | Temperature | Relative Humidity | Absolute Humidity |
| --- | --- | --- | --- |
| p-value | 0.020 | 0.019 | 0.014 |
| **No. of doctors** | | | |
| coef | -0.0109 | -0.0111 | -0.0111 |
| 95%CI | [-0.0167, -0.0051] | [-0.0167, -0.0054] | [-0.0168, -0.0053] |
| std.err | 0.0021 | 0.0020 | 0.0021 |
| t-stat | -5.21 | -5.45 | -5.37 |
| p-value | 0.006 | 0.006 | 0.006 |
| **Drop of BMI** | | | |
| coef | -0.5174 | -0.4236 | -0.5370 |
| 95%CI | [-0.8038, -0.2309] | [-0.6320, -0.2152] | [-0.8650, -0.2090] |
| std.err | 0.1032 | 0.0751 | 0.1181 |
| t-stat | -5.01 | -5.64 | -4.55 |
| p-value | 0.007 | 0.005 | 0.010 |
| **Inflow population from Wuhan** | | | |
| coef | -0.0006 | -0.0004 | -0.0005 |
| 95%CI | [-0.0010,-0.0001] | [-0.0009, 0.00003] | [-0.0010, -8.04E-6] |
| std.err | 0.0001 | 0.0002 | 0.0002 |
| t-stat | -3.70 | -2.57 | -2.82 |
| p-value | 0.021 | 0.062 | 0.048 |
| **Latitude** | | | |
| coef | 0.0283 | 0.0422 | 0.0396 |
| 95%CI | [0.0104, 0.0461] | [0.0331, 0.0512] | [0.0267, 0.0525] |
| std.err | 0.0064 | 0.0032 | 0.0046 |
| t-stat | 4.40 | 12.98 | 8.53 |
| p-value | 0.012 | 0.0002 | 0.001 |
| **Longitude** | | | |
| coef | -0.0299 | -0.0273 | -0.0289 |
| 95%CI | [-0.0559, -0.0039] | [-0.0523, -0.0023] | [-0.0543, -0.0034] |
| std.err | 0.0094 | 0.0090 | 0.0092 |
| t-stat | -3.19 | -3.03 | -3.15 |
| p-value | 0.033 | 0.039 | 0.035 |
| **const** | | | |
| coef | 2.1182 | 2.1184 | 2.1176 |
| 95%CI | [1.5681, 2.6684] | [1.5667, 2.6700] | [1.5682, 2.6670] |
| std.err | 0.1981 | 0.1987 | 0.1979 |
| t-stat | 10.69 | 10.66 | 10.70 |
| p-value | 0.0004 | 0.0004 | 0.0004 |

## Panel B: Regression for the U.S. Counties

|  | Temperature | Relative Humidity | Absolute Humidity |
|---|---|---|---|
| $R^2$ | 0.1210 | 0.1257 | 0.1255 |
| **Temperature** | | | |
| coef | -0.0138 | | |
| 95%CI | [-0.0267,-0.0009] | | |
| std.err | 0.0062 | | |
| t-stat | -2.21 | | |
| p-value | 0.038 | | |
| **Relative Humidity** | | | |
| coef | | -0.0078 | |
| 95%CI | | [-0.0140, -0.0014] | |
| std.err | | 0.0031 | |
| t-stat | | -2.53 | |
| p-value | | 0.019 | |
| **Absolute Humidity** | | | |
| coef | | | -0.0496 |
| 95%CI | | | [-0.0664, -0.0327] |
| std.err | | | 0.0081 |
| t-stat | | | -6.11 |
| p-value | | | 0 |
| **Population Density** | | | |
| coef | 6.51E-6 | 6.25E-6 | 5.50E-6 |
| 95%CI | [-0.00002, 0.00004] | [-0.00003, 0.00004] | [-0.00002, 0.00004] |
| std.err | 0.00002 | 0.00002 | 0.00001 |
| t-stat | 0.43 | 0.40 | 0.38 |
| p-value | 0.671 | 0.689 | 0.711 |
| **Percentage over 65** | | | |
| coef | -0.9306 | -1.0137 | -0.9071 |
| 95%CI | [-1.5574, -0.3038] | [-1.7090, -0.3183] | [-1.6107, -0.2034] |
| std.err | 0.3022 | 0.3353 | 0.339 |
| t-stat | -3.08 | -3.02 | -2.67 |
| p-value | 0.005 | 0.006 | 0.014 |
| **Gini** | | | |
| coef | -1.6920 | -1.8024 | -1.7177 |
| 95%CI | [-4.4260, 1.0420] | [-4.3390, 0.7342] | [-4.3598, 0.9263] |
| std.err | 1.3183 | 1.2231 | 1.2744 |
| t-stat | -1.28 | -1.47 | -1.35 |
| p-value | 0.213 | 0.155 | 0.192 |
| **Socio-economic factor** | | | |
| coef | 0.1371 | 0.1265 | 0.1363 |
| 95%CI | [0.0842, 0.1900] | [0.0783, 0.1747] | [0.0914, 0.1812] |
| std.err | 0.0255 | 0.0232 | 0.0217 |

|  | Temperature | Relative Humidity | Absolute Humidity |
| --- | --- | --- | --- |
| t-stat | 5.38 | 5.44 | 6.30 |
| p-value | 0.00002 | 0.00002 | 0 |
| **No. of ICU beds per capita** | | | |
| coef | -0.0122 | -0.0097 | -0.0127 |
| 95%CI | [-0.0359,0.0114] | [-0.0294,0.0100] | [-0.0351,-0.0097] |
| std.err | 0.0114 | 0.0095 | 0.0108 |
| t-stat | -1.07 | -1.02 | -1.17 |
| p-value | 0.294 | 0.317 | 0.253 |
| **Fraction of maximum moving distance over normal time** | | | |
| coef | 0.0005 | 0.0014 | 0.0011 |
| 95%CI | [-0.0038,0.0048] | [-0.0015, 0.0043] | [-0.0023,0.0045] |
| std.err | 0.0021 | 0.0014 | 0.0016 |
| t-stat | 0.24 | 0.98 | 0.65 |
| p-value | 0.815 | 0.338 | 0.520 |
| **Home stay minutes** | | | |
| coef | 0.0006 | 0.0006 | 0.0006 |
| 95%CI | [0.0003, 0.0009] | [0.0003,0.0010] | [0.0003, 0.0010] |
| std.err | 0.0001 | 0.0002 | 0.0002 |
| t-stat | 3.94 | 3.91 | 3.88 |
| p-value | 0.001 | 0.001 | 0.001 |
| **Latitude** | | | |
| coef | -0.0201 | -0.0097 | -0.0361 |
| 95%CI | [-0.0367, -0.0036] | [-0.0174, -0.0020] | [-0.0511, -0.0211] |
| std.err | 0.0080 | 0.0037 | 0.0072 |
| t-stat | -2.53 | -2.61 | -4.98 |
| p-value | 0.019 | 0.016 | 0.00006 |
| **Longitude** | | | |
| coef | 0.0104 | 0.0098 | 0.0107 |
| 95%CI | [0.0084, 0.0123] | [0.0079, 0.0117] | [0.0086,0.0128] |
| std.err | 0.0009 | 0.0009 | 0.0010 |
| t-stat | 11.02 | 10.66 | 10.52 |
| p-value | 0 | 0 | 0 |
| **const** | | | |
| coef | 2.2121 | 2.1911 | 2.2137 |
| 95%CI | [1.6662, 2.7580] | [1.6600, 2.7222] | [1.6659, 2.7616] |
| std.err | 0.2632 | 0.2561 | 0.2641 |
| t-stat | 8.40 | 8.56 | 8.38 |
| p-value | 0 | 0 | 0 |

# Supplementary Materials for

Impact of Temperature and Relative Humidity on the Transmission of COVID-19: A Modeling Study in China and the U.S.

Jingyuan Wang, Ke Tang[*], Kai Feng, Xin Lin, Weifeng Lv, Kun Chen and Fei Wang

[*]Correspondence to: ketang@tsinghua.edu.cn

**This PDF file includes:**

    Materials and Methods
    Figs. S1
    Tables S1 to S11

**Materials and Methods**

**Fama-MacBeth Regression with Newey-West Adjustment**

Fama-MacBeth regression is a way to study the relationship between the response variable and the features in the panel data setup. Particularly, Fama-MacBeth regression runs a series of cross-sectional regressions and uses the average of the cross-sectional regression coefficients as the second step of parameter estimation. In equation form, for $n$ response variables, $m$ features and time series length $T$

$$R_{i,1} = \alpha_1 + \beta_{1,1} F_{1,i,1} + \beta_{2,1} F_{2,i,1} + \cdots + \beta_{m,1} F_{m,i,1} + \epsilon_{i,1},$$
$$R_{i,2} = \alpha_2 + \beta_{1,2} F_{1,i,2} + \beta_{2,2} F_{2,i,2} + \cdots + \beta_{m,2} F_{m,i,2} + \epsilon_{i,2},$$
$$\cdots$$
$$R_{i,T} = \alpha_T + \beta_{1,T} F_{1,i,T} + \beta_{2,T} F_{2,i,T} + \cdots + \beta_{m,T} F_{m,i,T} + \epsilon_{i,T}.$$

where $R_{i,t}, i \in \{1, \ldots, n\}$ are the response values, $\beta_{k,t}$ are first step regression coefficients for feature $k$ at time $t$, and $F_{k,i,t}$ are the input features of feature $k$ and sample $i$ at time $t$. In the second step, the average of the first step regression coefficient, $\hat{\beta}_k$, can be calculated directly, or via the following regression

$$\beta_{k,t} = c_k + \epsilon_t.$$

where $\epsilon_t$ is the random noise.

Since $\beta$s might have time-series autocorrelation, in the second step, we thus use the Newey-West approach [1] to adjust the time-series autocorrelation (and heteroscedasticity) in calculating standard errors. Specifically, for the second step, we have $E[\epsilon] = 0$ and $E[\epsilon \epsilon'] = \sigma^2 \Omega$.

The covariance matrix of $c_k$ is

$$V_{c_k} = \frac{1}{T} \left(\frac{1}{T} \mathbf{1}' \mathbf{1}\right)^{-1} \left(\frac{1}{T} \mathbf{1}'(\sigma^2 \Omega) \mathbf{1}\right) \left(\frac{1}{T} \mathbf{1}' \mathbf{1}\right)^{-1},$$

where $\mathbf{1}$ is a $T \times 1$ vector of 1 and $\sigma^2 \Omega$ is the covariance matrix of errors.

The middle matrix can be rewritten as

$$Q = \frac{1}{T}\mathbf{1}'(\sigma^2 \Omega)\mathbf{1}$$
$$= \frac{1}{T}\sum_{i=1}^{T}\sum_{j=1}^{T}\sigma_{ij}$$

The Newey-West estimators give a consistent estimation of $Q$ when the residuals are autocorrelated and/or heteroscedastic. The Newey-West estimator can be expressed as

$$S = \frac{1}{T}\left(\sum_{t=1}^{T}e_t^2 + \sum_{l=1}^{L}\sum_{t=l+1}^{T}w_l e_t e_{t-l}\right),$$

where $w_l = 1 - \frac{l}{1+L}$, e represents residuals and $L$ is the lag.

We use Fama-Macbeth regressions for two reasons. First, the temperature and relative humidity series have trends with the arrival of summer and the $R$ value series also has downward trends. In this case, panel regression will obtain spurious regression results from the time-series perspective. However, the cross-sectional regression involving cities (counties) of various meteorological conditions and COVID-19 spread intensities will not have spurious regression issues. Second, Fama-MacBeth regression is valid even in the presence of cross-sectional heteroskedasticity (including complex spatial covariance) because in the second-step regression, only the value of the first step estimates $\beta$s are used, not their standard errors. Therefore, as long as the first-step estimator is unbiased, which is the case for heteroskedasticity (including complex spatial covariance), the Fama-MacBeth estimation is correct.

Less rigorously speaking, we use the first step of Fama-MacBeth regression to determine the extent to which the transmissibility of the areas of high temperature and high relative humidity are compared with that of low temperature and low relative humidity areas each day. We then use the second step to test whether daily relationships are a common fact during a given time period.

**Estimating the Effective Reproduction Number**

The basic reproduction number $R_0$, which characterizes the transmission ability of an epidemic, is defined as the average number of people who will contract the contagious disease from a typical infected case in a population where everyone is susceptible. When an epidemic spreads through a population, the time-varying effective reproduction number $R_t$ is of greater concern. The effective reproduction number $R_t$, the $R$ value at time step $t$, is defined as the actual average number of secondary cases per primary case cause[2].

We then calculate the effective reproductive number $R_t$ for each city through a time-dependent method based on maximun likelihood estimation (MLE)[3]. The inputs to the method are epidemic curves, *i.e.*, the historical numbers of patients in each day, for a certain city. Specifically, we denote $w(\tau|\theta)$ as the probability distribution for the serial interval, which is defined as the time between symptom onset of a case and symptom onset of her/his secondary cases. Let $p_{(i,j)}$ be the relative likelihood that case *i* has been infected by case *j*, given the difference in time of symptom onset $t_i - t_j$, which can be expressed in terms of $w(\tau|\theta)$. That is, the relative likelihood that case *i* has been infected by case *j* can be expressed as

$$p_{ij} = \frac{w(t_i - t_j)}{\sum_{i \neq k} w(t_i - t_k)}$$

The relative likelihood of case *i* infecting case *j* is independent of the relative likelihood of case *i* infecting any other case *k*. The distribution of the effective reproduction number for case *i* is

$$R_i \sim \sum_j \text{Bernoulli}[p_{(j,i)}]$$

With the expected value

$$E(R_i) = \sum_j p_{(j,i)}$$

The average daily effective reproduction number $R_t$ is estimated as the average over $R_i$ for all cases *i* who develop the first symptom of onset on day *t*.

The above calculation is implemented with the package 'R0' developed by Boelle & Obadia with R version 3.6.2 and 'R0' version 1.2_6 (https://cran.r-project.org/web/packages/R0/index.html).

**Modeling Spatial Effect**

We use a generalized linear mixed model (GLMM) with spatial random effects to account for spatial autocorrelation between cities or counties in each cross-sectional regression. The form of the model is

$$\boldsymbol{y} = \boldsymbol{X\beta} + \boldsymbol{u} + \boldsymbol{\epsilon},$$

where $\boldsymbol{y}$ is the $N \times 1$ outcome vector, $\boldsymbol{X}$ is the $N \times p$ matrix of the $p$ explanatory variables (the intercept term can be included by setting the first column of X as a vector of ones), $\boldsymbol{\beta}$ is the vector of regression coefficients, $\boldsymbol{u}$ is the vector of spatial random effects, and $\boldsymbol{\epsilon}$ is the random error vector whose entries are independent and identically distributed as $N(0, \sigma^2)$. We assume $\boldsymbol{u} \sim N(0, \boldsymbol{\sigma_s^2 G})$, where $\boldsymbol{\sigma_s^2}$ is the spatial variance and $\boldsymbol{G}$ follows a Matérn correlation structure[4].

The Matérn model flexibly specifies the correlation between any two cities or counties as a function of their geographical distance; the model has two parameters, a scale parameter $\rho$ and a "smoothness" parameter $\nu$, and it subsumes the exponential and squared exponential models as special cases. The maximum likelihood method is used for parameter estimation[5].

We have also tried a conditional autoregressive model (CAR)[6] in which the spatial correlation is described by an adjacency matrix of the cities/counties. The Matérn model performs better than the CAR model as judged by the Akaike information criterion (AIC); the average AIC value across all cross-sectional regressions is 896.9 and 936.5 for the Matérn model and the CAR model, respectively.

All computations are performed in the R package "spaMM" version 3.3.0[7]. We report the results from the Matérn model in Table S9 and S10.

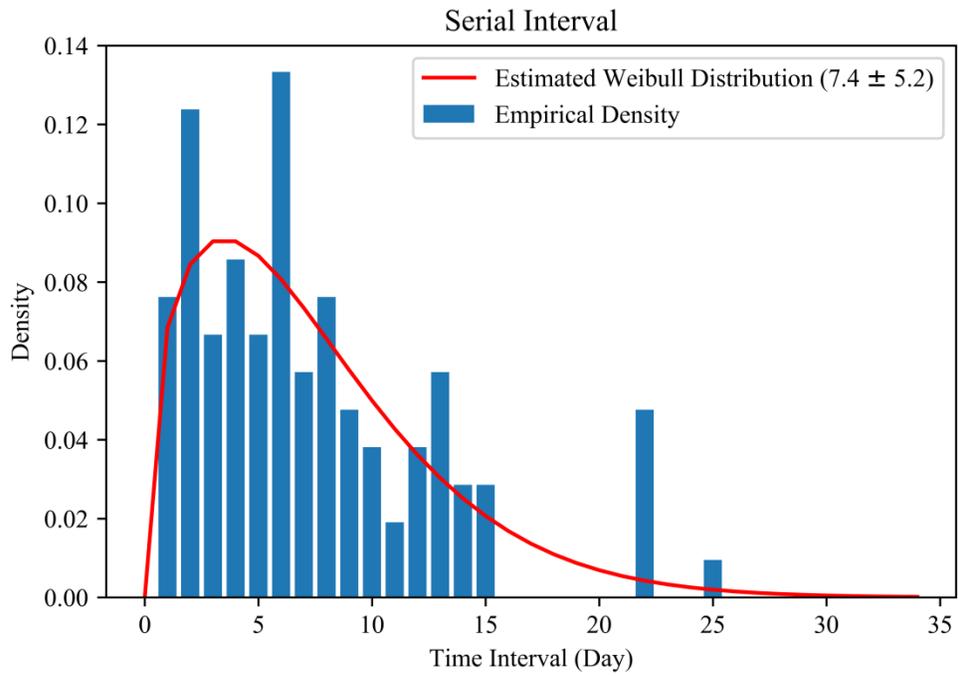

**Fig. S1. Estimation of the serial interval with the Weibull distribution**

Bars denote the probability of occurrences in specified bins, and the red curve is the density function of the estimated Weibull distribution.

**Table S1. Data Summary**

This table summarizes the variables used in this paper. Panel A and B summarize the data of Chinese cities and the U.S. counties.

| Panel A: Data Summary for the Chinese Cities | | | | |
|---|---|---|---|---|
| | Mean | Std | Min | Max |
| R | 1.072 | 0.707 | 0.131 | 4.609 |
| 6-Day Average Temperature (Celsius) | 4.468 | 6.842 | -21.100 | 19.733 |
| 6-Day Average Relative Humidity (%) | 77.147 | 9.589 | 48.667 | 99.833 |
| GDP per Capita (RMB 10k) | 6.800 | 3.716 | 2.159 | 18.957 |
| Population Density (k/km$^2$) | 0.692 | 0.812 | 0.00800 | 6.522 |
| No. Doctors (k) | 16.020 | 11.488 | 1.972 | 68.549 |
| Proxy for Inflow population from Wuhan (10 k) | 5.096 | 14.833 | 0.000 | 138.154 |
| Fraction over 65 | 0.121 | 0.0186 | 0.0826 | 0.152 |
| Drop of BMI compared to first week 2020 | -0.413 | 0.347 | -0.886 | 0.759 |

| Panel B: Data Summary for the U.S. Counties | | | | |
|---|---|---|---|---|
| | Mean | Std | Min | Max |
| R | 1.517 | 0.836 | 0.040 | 4.997 |
| 6-Day Average Temperature (Celsius) | 10.738 | 6.503 | -10.192 | 28.826 |
| 6-Day Average Relative Humidity (%) | 67.815 | 11.932 | 16.388 | 99.096 |
| Population Density (/mile$^2$) | 374.275 | 1678.13 | 2.562 | 48229.375 |
| Fraction over 65 | 0.167 | 0.0423 | 0.0633 | 0.374 |
| Gini index | 0.449 | 0.0309 | 0.357 | 0.597 |
| GDP per capita (k Dollar) | 45.599 | 24.417 | 13.006 | 378.762 |
| Fraction below poverty level | 15.970 | 5.604 | 4.000 | 38.100 |
| Personal income (Dollar) | 46923.2 | 14586.7 | 26407 | 251728 |
| Fraction of not in labor force, 16 years or over | 38.842 | 6.737 | 19.600 | 62.000 |
| Fraction of total household more than $200,000 | 3.564 | 2.948 | 0.400 | 23.100 |
| Fraction of food stamp/SNAP benefits | 13.854 | 5.355 | 1.400 | 38.800 |
| No. ICU beds per 10000 capita | 2.182 | 1.945 | 0.000 | 17.357 |
| Fraction of maximum moving distance over normal time | 33.286 | 25.918 | 0.000 | 478.000 |
| Home-stay minutes | 749.064 | 145.883 | 206.585 | 1275.341 |

**Table S2: Pairwise Correlation Analysis for Chinese Cities**

Pairwise correlation coefficients are obtained by averaging all correlation coefficients from each time step in the Fama-Macbeth approach.

|  | Temperature | Relative Humidity | Population Density | Percentage over 65 | GDP per capita | No. of doctors | Drop of BMI | Inflow population from Wuhan | Latitude | Longitude |
|---|---|---|---|---|---|---|---|---|---|---|
| Temperature | 1.00 | 0.32 | 0.33 | -0.37 | 0.33 | 0.13 | -0.21 | 0.04 | -0.92 | -0.57 |
| Relative Humidity | 0.32 | 1.00 | -0.08 | 0.01 | -0.16 | -0.09 | 0.29 | 0.09 | -0.44 | -0.32 |
| Population Density | 0.33 | -0.08 | 1.00 | -0.27 | 0.57 | 0.29 | -0.40 | -0.09 | -0.27 | -0.03 |
| Percentage over 65 | -0.37 | 0.01 | -0.27 | 1.00 | -0.20 | 0.13 | 0.25 | 0.06 | 0.45 | 0.13 |
| GDP per capita | 0.33 | -0.16 | 0.57 | -0.20 | 1.00 | 0.45 | -0.76 | -0.14 | -0.25 | 0.05 |
| No. of doctors | 0.13 | -0.09 | 0.29 | 0.13 | 0.45 | 1.00 | -0.39 | -0.12 | -0.06 | -0.22 |
| Drop of BMI | -0.21 | 0.29 | -0.40 | 0.25 | -0.76 | -0.39 | 1.00 | 0.04 | 0.12 | -0.14 |
| Inflow population from Wuhan | 0.04 | 0.09 | -0.09 | 0.06 | -0.14 | -0.12 | 0.04 | 1.00 | -0.05 | -0.12 |
| Latitude | -0.92 | -0.44 | -0.27 | 0.45 | -0.25 | -0.06 | 0.12 | -0.05 | 1.00 | 0.59 |
| Longitude | -0.57 | -0.32 | -0.03 | 0.13 | 0.05 | -0.22 | -0.14 | -0.12 | 0.59 | 1.00 |

## Table S3: Pairwise Correlation Analysis for the U.S. Counties

Pairwise correlation coefficients are obtained by averaging all correlation coefficients from each time step in the Fama-Macbeth approach.

|  | Temperature | Relative Humidity | Population Density | Percentage over 65 | Gini | Se-factor | No. of ICU beds per capita | M50_index | Home stay minutes | Latitude | Longitude |
|---|---|---|---|---|---|---|---|---|---|---|---|
| Temperature | 1.00 | 0.17 | 0.01 | -0.05 | 0.34 | 0.36 | 0.11 | 0.34 | 0.00 | -0.90 | 0.04 |
| Relative Humidity | 0.17 | 1.00 | -0.06 | 0.08 | 0.05 | 0.02 | 0.00 | 0.07 | 0.10 | -0.20 | 0.12 |
| Population Density | 0.01 | -0.06 | 1.00 | -0.11 | 0.23 | 0.07 | 0.07 | -0.19 | 0.11 | 0.01 | 0.10 |
| Percentage over 65 | -0.05 | 0.08 | -0.11 | 1.00 | 0.02 | 0.14 | -0.04 | -0.03 | -0.18 | 0.05 | 0.13 |
| Gini | 0.34 | 0.05 | 0.23 | 0.02 | 1.00 | 0.53 | 0.37 | 0.15 | -0.17 | -0.35 | 0.07 |
| Socio-economic factor | 0.36 | 0.02 | 0.07 | 0.14 | 0.53 | 1.00 | 0.21 | 0.32 | -0.41 | -0.34 | 0.00 |
| No. of ICU beds per capita | 0.11 | 0.00 | 0.07 | -0.04 | 0.37 | 0.21 | 1.00 | 0.18 | -0.10 | -0.11 | 0.10 |
| M50_index | 0.34 | 0.07 | -0.19 | -0.03 | 0.15 | 0.32 | 0.18 | 1.00 | -0.37 | -0.37 | -0.08 |
| Home-stay minutes | 0.00 | 0.10 | 0.11 | -0.18 | -0.17 | -0.41 | -0.10 | -0.37 | 1.00 | 0.06 | -0.08 |
| Latitude | -0.90 | -0.20 | 0.01 | 0.05 | -0.35 | -0.34 | -0.11 | -0.37 | 0.06 | 1.00 | -0.06 |
| Longitude | 0.04 | 0.12 | 0.10 | 0.13 | 0.07 | 0.00 | 0.10 | -0.08 | -0.08 | -0.06 | 1.00 |

**Table S4: Unit Root Test for R, Temperature and Relative Humidity**

Panel A and B show the results of Handri LM test [8] with null hypotheses of non-unit-roots, for Chinese cities and the U.S. counties, respectively.

|  | Panel A: Test Results for Chinese Cities | | |
| --- | --- | --- | --- |
|  | *R* value | Temperature | Relative Humidity |
| **z-stat** | 18.7472 | 51.1532 | 42.6092 |
| **p-value** | 0.0000 | 0.0000 | 0.0000 |
|  | Panel B: Test Results for the U.S. Counties | | |
|  | *R* value | Temperature | Relative Humidity |
| **z-stat** | 43.0116 | 61.0510 | 76.8665 |
| **p-value** | 0.0000 | 0.0000 | 0.0000 |

**Table S5: Coefficients of temperature and relative humidity in first step of Fama-Macbeth Regression**

Panel A and B show regression coefficients of temperature and relative humidity in the first step of Fama-Macbeth regression, for Chinese cities and the U.S. counties, respectively.

| Panel A: Regression Coefficients for Chinese Cities | | |
|---|---|---|
| Date | Coefficient of Temperature | Coefficient of Relative Humidity |
| Jan, 19 | -0.0373 | -0.0109 |
| Jan, 20 | -0.0064 | 0.0009 |
| Jan, 21 | -0.0127 | -0.0093 |
| Jan, 22 | -0.0309 | -0.0121 |
| Jan, 23 | -0.0427 | -0.0066 |
| Jan, 24 | -0.0249 | 0.0010 |
| Jan, 25 | -0.0238 | -0.0062 |
| Jan, 26 | -0.0506 | -0.0174 |
| Jan, 27 | -0.0526 | -0.0159 |
| Jan, 28 | -0.0196 | -0.0063 |
| Jan, 29 | -0.0340 | -0.0101 |
| Jan, 30 | -0.0305 | -0.0096 |
| Jan, 31 | -0.0391 | -0.0087 |
| Feb, 1 | -0.0388 | -0.0102 |
| Feb, 2 | -0.0248 | -0.0097 |
| Feb, 3 | -0.0108 | -0.0022 |
| Feb, 4 | -0.0091 | 0.0020 |
| Feb, 5 | 0.0039 | 0.0040 |
| Feb, 6 | -0.0061 | -0.0037 |
| Feb, 7 | -0.0034 | 0.0006 |
| Feb, 8 | 0.0103 | -0.0030 |
| Feb, 9 | -0.0077 | -0.0067 |
| Feb, 10 | -0.0150 | 0.0052 |

| Panel B: Regression Coefficients for U.S. Counties | | |
|---|---|---|
| Date | Coefficient of Temperature | Coefficient of Relative Humidity |
| **Mar, 15** | -0.0402 | -0.0190 |
| **Mar, 16** | -0.0309 | -0.0192 |
| **Mar, 17** | -0.0052 | -0.0129 |
| **Mar, 18** | -0.0192 | -0.0146 |
| **Mar, 19** | -0.0412 | -0.0237 |
| **Mar, 20** | 0.0224 | -0.0114 |
| **Mar, 21** | -0.0112 | -0.0158 |
| **Mar, 22** | -0.0138 | -0.0169 |
| **Mar, 23** | -0.0021 | -0.0195 |
| **Mar, 24** | -0.0107 | -0.0166 |
| **Mar, 25** | -0.0184 | -0.0073 |
| **Mar, 26** | -0.0231 | -0.0095 |
| **Mar, 27** | -0.0241 | -0.0010 |
| **Mar, 28** | -0.0468 | 0.0013 |
| **Mar, 29** | -0.0314 | 0.0007 |
| **Mar, 30** | -0.0533 | 0.0076 |
| **Mar, 31** | -0.0403 | 0.0071 |
| **Apr, 1** | -0.0386 | -0.0003 |
| **Apr, 2** | -0.0234 | -0.0017 |
| **Apr, 3** | 0.0029 | -0.0024 |
| **Apr, 4** | 0.0037 | -0.0031 |
| **Apr, 5** | -0.0177 | -0.0010 |
| **Apr, 6** | -0.0057 | -0.0040 |
| **Apr, 7** | -0.0041 | -0.0028 |
| **Apr, 8** | -0.0116 | -0.0029 |
| **Apr, 9** | -0.0138 | -0.0032 |
| **Apr, 10** | -0.0123 | -0.0032 |
| **Apr, 11** | -0.0211 | -0.0021 |

| Date | Coefficient of Temperature | Coefficient of Relative Humidity |
|---|---|---|
| Apr, 12 | -0.0297 | -0.0002 |
| Apr, 13 | -0.0244 | -0.0008 |
| Apr, 14 | -0.0310 | -0.0016 |
| Apr, 15 | -0.0295 | -0.0012 |
| Apr, 16 | -0.0271 | -0.0010 |
| Apr, 17 | -0.0297 | 0.0022 |
| Apr, 18 | -0.0245 | 0.0027 |
| Apr, 19 | -0.0196 | 0.0020 |
| Apr, 20 | -0.0110 | -0.0012 |
| Apr, 21 | 0.0068 | -0.0002 |
| Apr, 22 | 0.0126 | -0.0015 |
| Apr, 23 | 0.0061 | -0.0033 |
| Apr, 24 | 0.0216 | -0.0028 |
| Apr, 25 | 0.0186 | -0.0030 |

**Table S6: Fama-Macbeth Regression for Chinese Cities except Wuhan**

Daily $R$ values from January 19 to February 10 and the average temperature and relative humidity over 6 days up to and including the day when $R$ value is measured, are used in the regression for 99 Chinese cities (without Wuhan). The regression is estimated by the Fama-MacBeth approach.

|  | Overall | Before Lockdown (Jan 24) | After Lockdown (Jan 24) |
|---|---|---|---|
| R2 | 0.3029 | 0.1915 | 0.3339 |
| **Temperature** | | | |
| coef | -0.0223 | -0.0287 | -0.0205 |
| 95%CI | [-0.0358, -0.0088] | [-0.0406, -0.0168] | [-0.0369, -0.0041] |
| std.err | 0.0065 | 0.0043 | 0.0078 |
| t-stat | -3.44 | -6.69 | -2.64 |
| p-value | 0.002 | 0.003 | 0.017 |
| **Relative Humidity** | | | |
| coef | -0.0060 | -0.0071 | -0.0056 |
| 95%CI | [-0.0100, -0.0019] | [-0.0105, -0.0038] | [-0.0108, -0.0005] |
| std.err | 0.0019 | 0.0012 | 0.0024 |
| t-stat | -3.07 | -5.86 | -2.32 |
| p-value | 0.006 | 0.004 | 0.033 |
| **Population Density** | | | |
| coef | 0.0262 | 0.1198 | 0.0002 |
| 95%CI | [-0.0290, 0.0814] | [0.0564, 0.1832] | [-0.0352, 0.0356] |
| std.err | 0.0266 | 0.0228 | 0.0168 |
| t-stat | 0.98 | 5.25 | 0.01 |
| p-value | 0.336 | 0.006 | 0.991 |
| **Percentage over 65** | | | |
| coef | 0.1316 | 0.3849 | 0.0612 |
| 95%CI | [-1.7302, 1.9933] | [-1.0386, 1.8084] | [-2.3111, 2.4335] |
| std.err | 0.8977 | 0.5127 | 1.1244 |
| t-stat | 0.15 | 0.75 | 0.05 |

|  | Overall | Before Lockdown (Jan 24) | After Lockdown (Jan 24) |
| --- | --- | --- | --- |
| p-value | 0.885 | 0.495 | 0.957 |
| **GDP per capita** | | | |
| coef | 0.0048 | -0.0110 | 0.0092 |
| 95%CI | [-0.0148, 0.0244] | [-0.0252, 0.0033] | [-0.0114, 0.0298] |
| std.err | 0.0095 | 0.0051 | 0.0098 |
| t-stat | 0.51 | -2.13 | 0.94 |
| p-value | 0.616 | 0.100 | 0.360 |
| **No. of doctors** | | | |
| coef | -0.0057 | -0.0109 | -0.0043 |
| 95%CI | [-0.0089, -0.0025] | [-0.0162, -0.0056] | [-0.0064, -0.0022] |
| std.err | 0.0015 | 0.0019 | 0.0010 |
| t-stat | -3.73 | -5.69 | -4.35 |
| p-value | 0.001 | 0.005 | 0.0004 |
| **Drop of BMI** | | | |
| coef | 0.3135 | -0.4107 | 0.5146 |
| 95%CI | [-0.3290, -0.9559] | [-0.6870, -0.1344] | [-0.0995, 1.1287] |
| std.err | 0.3098 | 0.0995 | 0.2911 |
| t-stat | 1.01 | -4.13 | 1.77 |
| p-value | 0.323 | 0.015 | 0.095 |
| **Inflow population from Wuhan** | | | |
| coef | -0.0052 | -0.0006 | -0.0065 |
| 95%CI | [-0.0106, 0.0002] | [-0.0011, -0.0002] | [-0.0128, -0.0002] |
| std.err | 0.0026 | 0.0002 | 0.0030 |
| t-stat | -1.99 | -3.93 | -2.17 |
| p-value | 0.059 | 0.017 | 0.044 |
| **Latitude** | | | |
| coef | 0.0040 | 0.0082 | 0.0029 |
| 95%CI | [-0.0149, 0.0230] | [-0.0132, 0.0296] | [-0.0213, 0.0271] |
| std.err | 0.0091 | 0.0077 | 0.0115 |

|  | Overall | Before Lockdown (Jan 24) | After Lockdown (Jan 24) |
|---|---|---|---|
| t-stat | 0.44 | 1.06 | 0.25 |
| p-value | 0.663 | 0.347 | 0.804 |
| **Longitude** | | | |
| coef | -0.0110 | -0.0293 | -0.0059 |
| 95%CI | [-0.0209, -0.0010] | [-0.0579, -0.0008] | [-0.0134, 0.0017] |
| std.err | 0.0048 | 0.0103 | 0.0036 |
| t-stat | -2.29 | -2.85 | -1.64 |
| p-value | 0.032 | 0.046 | 0.119 |
| **const** | | | |
| coef | 1.0925 | 2.1209 | 0.8069 |
| 95%CI | [0.5059, 1.6792] | [1.5697, 2.6721] | [0.5327, 1.0810] |
| std.err | 0.2829 | 0.1985 | 0.1299 |
| t-stat | 3.86 | 10.68 | 6.21 |
| p-value | 0.001 | 0 | 0 |

**Table S7: Relationship between Temperature, Relative Humidity, and *R* Values: Robustness Check with the Serial Interval of Mean 7.5 Days and Standard Deviation 3.4 days in Li et al (2020)[2] for Chinese Cities**

This table utilizes the estimated serial interval in a previous paper (mean 7.5 days, std 3.4 days)[2] to construct *R* values for China. The table reports the coefficients of the effective reproductive number, *R* values, on an intercept, temperature, relative humidity and control variables in the Fama-MacBeth regressions.

|  | Overall | Before Lockdown (Jan 24) | After Lockdown (Jan 24) |
|---|---|---|---|
| R2 | 0.2843 | 0.2009 | 0.3074 |
| **Temperature** | | | |
| coef | -0.0267 | -0.0430 | -0.0222 |
| 95%CI | [-0.0486,-0.0048] | [-0.0694,-0.0165] | [-0.0456,0.0012] |
| std.err | 0.0106 | 0.0095 | 0.0111 |
| t-stat | -2.53 | -4.52 | -2.00 |
| p-value | 0.019 | 0.011 | 0.061 |
| **Relative Humidity** | | | |
| coef | -0.0076 | -0.0104 | -0.0068 |
| 95%CI | [-0.0121,-0.0031] | [-0.0166,-0.0041] | [-0.0121,-0.0015] |
| std.err | 0.0022 | 0.0023 | 0.0025 |
| t-stat | -3.47 | -4.59 | -2.69 |
| p-value | 0.002 | 0.010 | 0.015 |
| **Population Density** | | | |
| coef | 0.0223 | 0.1673 | -0.0180 |
| 95%CI | [-0.0672,0.1118] | [0.0350,0.2996] | [-0.0825,0.0465] |
| std.err | 0.0432 | 0.0477 | 0.0306 |
| t-stat | 0.52 | 3.51 | -0.59 |
| p-value | 0.611 | 0.025 | 0.563 |
| **Percentage over 65** | | | |
| coef | -0.7581 | 0.3976 | -1.0791 |

|  | Overall | Before Lockdown (Jan 24) | After Lockdown (Jan 24) |
| --- | --- | --- | --- |
| 95%CI | [-3.7515,2.2353] | [-2.9474,3.7426] | [-4.8094,2.6511] |
| std.err | 1.4434 | 1.2048 | 1.7680 |
| t-stat | -0.53 | 0.33 | -0.61 |
| p-value | 0.605 | 0.758 | 0.550 |
| **GDP per capita** | | | |
| coef | 0.0058 | -0.0291 | 0.0154 |
| 95%CI | [-0.0246,0.0361] | [-0.0390,-0.0193] | [-0.0124,0.0433] |
| std.err | 0.0147 | 0.0035 | 0.0132 |
| t-stat | 0.39 | -8.21 | 1.17 |
| p-value | 0.698 | 0.001 | 0.258 |
| **No. of doctors** | | | |
| coef | -0.0065 | -0.0135 | -0.0045 |
| 95%CI | [-0.0107,-0.0023] | [-0.0205,-0.0065] | [-0.0067,-0.0024] |
| std.err | 0.0020 | 0.0025 | 0.0010 |
| t-stat | -3.22 | -5.35 | -4.47 |
| p-value | 0.004 | 0.006 | 0.0003 |
| **Drop of BMI** | | | |
| coef | 0.3287 | -0.7465 | 0.6274 |
| 95%CI | [-0.5135,1.1709] | [-1.3448,-0.1483] | [-0.1037,1.3585] |
| std.err | 0.4061 | 0.2155 | 0.3465 |
| t-stat | 0.81 | -3.46 | 1.81 |
| p-value | 0.427 | 0.026 | 0.088 |
| **Inflow population from Wuhan** | | | |
| coef | -0.0053 | -0.0003 | -0.0067 |
| 95%CI | [-0.0114,0.0008] | [-0.0009,0.0003] | [-0.0139,0.0006] |
| std.err | 0.0029 | 0.0002 | 0.0034 |
| t-stat | -1.79 | -1.34 | -1.94 |
| p-value | 0.087 | 0.250 | 0.069 |
| **Latitude** | | | |

|  | Overall | Before Lockdown (Jan 24) | After Lockdown (Jan 24) |
| --- | --- | --- | --- |
| coef | 0.0026 | 0.0045 | 0.0021 |
| 95%CI | [-0.0245,0.0298] | [-0.0518,0.0608] | [-0.0302,0.0344] |
| std.err | 0.0131 | 0.0203 | 0.0153 |
| t-stat | 0.20 | 0.22 | 0.14 |
| p-value | 0.843 | 0.835 | 0.893 |
| **Longitude** | | | |
| coef | -0.0103 | -0.0305 | -0.0046 |
| 95%CI | [-0.0233,0.0027] | [-0.0796,0.0186] | [-0.0160,0.0067] |
| std.err | 0.0063 | 0.0177 | 0.0054 |
| t-stat | -1.64 | -1.72 | -0.86 |
| p-value | 0.116 | 0.16 | 0.399 |
| **const** | | | |
| coef | 1.0616 | 2.2036 | 0.7444 |
| 95%CI | [0.4353,1.6879] | [1.431,2.9762] | [0.5063,0.9826] |
| std.err | 0.3020 | 0.2783 | 0.1129 |
| t-stat | 3.52 | 7.92 | 6.60 |
| p-value | 0.002 | 0.001 | 0 |

**Table S8: Relationship between Temperature, Relative Humidity, and *R* Value: Robustness Check with the Serial Interval of Mean 7.5 Days and Standard Deviation 3.4 days in Li et al (2020)[2] for the U.S. Counties**

This table utilizes the estimated serial interval in a previous paper (mean 7.5 days, std 3.4 days)[2] to construct *R* values for the U.S. counties. The table reports the coefficients of the effective reproductive number, *R* value, on an intercept, temperature, relative humidity and control variables in the Fama-MacBeth regressions.

|  | Overall | Before Lockdown (April 7) | After Lockdown (April 7) |
|---|---|---|---|
| R2 | 0.1170 | 0.1508 | 0.0760 |
| **Temperature** | | | |
| coef | -0.0199 | -0.0271 | -0.0113 |
| 95%CI | [-0.0330,-0.0069] | [-0.0456,-0.0086] | [-0.0296,0.0071] |
| std.err | 0.0065 | 0.0089 | 0.0087 |
| t-stat | -3.08 | -3.03 | -1.29 |
| p-value | 0.004 | 0.006 | 0.214 |
| **Relative Humidity** | | | |
| coef | -0.0052 | -0.0086 | -0.0011 |
| 95%CI | [-0.0114,0.0011] | [-0.0169,-0.0003] | [-0.0030,0.0008] |
| std.err | 0.0031 | 0.0040 | 0.0009 |
| t-stat | -1.68 | -2.14 | -1.20 |
| p-value | 0.101 | 0.044 | 0.244 |
| **Population Density** | | | |
| coef | 0.00002 | 3.00E-05 | 5.07E-08 |
| 95%CI | [-0.00003,0.00006] | [-0.0001,0.0001] | [-2.20e-6,2.30e-6] |
| std.err | 0.00002 | 4.00E-05 | 1.07E-06 |
| t-stat | 0.73 | 0.71 | 0.05 |
| p-value | 0.469 | 0.483 | 0.963 |
| **Percentage over 65** | | | |
| coef | -0.9733 | -1.2685 | -0.6159 |

|  | Overall | Before Lockdown (April 7) | After Lockdown (April 7) |
|---|---|---|---|
| 95%CI | [-1.4465,-0.5000] | [-1.9245,-0.6124] | [-1.0408,-0.1911] |
| std.err | 0.2343 | 0.3163 | 0.2022 |
| t-stat | -4.15 | -4.01 | -3.05 |
| p-value | 0.0002 | 0.001 | 0.007 |
| **Gini** | | | |
| coef | -1.9913 | -2.4119 | -1.4822 |
| 95%CI | [-3.6305,-0.3521] | [-4.9880,0.1643] | [-2.2360,-0.7285] |
| std.err | 0.8117 | 1.2422 | 0.3588 |
| t-stat | -2.45 | -1.94 | -4.13 |
| p-value | 0.018 | 0.065 | 0.001 |
| **Socio-economic factor** | | | |
| coef | 0.0906 | 0.1424 | 0.0279 |
| 95%CI | [0.0166,0.1646] | [0.0627,0.2222] | [-0.0112,0.0670] |
| std.err | 0.0366 | 0.0385 | 0.0186 |
| t-stat | 2.47 | 3.70 | 1.50 |
| p-value | 0.018 | 0.001 | 0.152 |
| **No. of ICU beds per capita** | | | |
| coef | -0.0113 | -0.0127 | -0.0096 |
| 95%CI | [-0.0263,0.0038] | [-0.0367,0.0113] | [-0.0147,-0.0044] |
| std.err | 0.0075 | 0.0116 | 0.0025 |
| t-stat | -1.51 | -1.10 | -3.91 |
| p-value | 0.138 | 0.285 | 0.001 |
| **Fraction of maximum moving distance over normal time** | | | |
| coef | 0.0036 | 0.0019 | 0.0056 |
| 95%CI | [0.0006,0.0066] | [-0.0023,0.0061] | [0.0043,0.0070] |
| std.err | 0.0015 | 0.0020 | 0.0007 |
| t-stat | 2.44 | 0.94 | 8.67 |
| p-value | 0.019 | 0.356 | 0 |
| **Home-stay minutes** | | | |

|  | **Overall** | **Before Lockdown (April 7)** | **After Lockdown (April 7)** |
|---|---|---|---|
| coef | 0.0003 | 0.0007 | -0.0003 |
| 95%CI | [-0.0003,0.0008] | [0.0003,0.0011] | [-0.0005,-2e-05] |
| std.err | 0.0003 | 0.0002 | 0.0001 |
| t-stat | 1.00 | 3.28 | -2.24 |
| p-value | 0.321 | 0.003 | 0.038 |
| **Latitude** | | | |
| coef | -0.0259 | -0.0514 | 0.0049 |
| 95%CI | [-0.0551,0.0032] | [-0.0825,-0.0203] | [-0.0179,0.0277] |
| std.err | 0.0144 | 0.0150 | 0.0109 |
| t-stat | -1.80 | -3.43 | 0.45 |
| p-value | 0.080 | 0.002 | 0.657 |
| **Longitude** | | | |
| coef | 0.0070 | 0.0110 | 0.0021 |
| 95%CI | [0.0019,0.0120] | [0.0059,0.0161] | [0.0003,0.0039] |
| std.err | 0.0025 | 0.0025 | 0.0009 |
| t-stat | 2.79 | 4.45 | 2.50 |
| p-value | 0.008 | 0.0002 | 0.022 |
| **const** | | | |
| coef | 1.7601 | 2.2325 | 1.1882 |
| 95%CI | [1.1636,2.3566] | [1.6514,2.8137] | [1.1588,1.2177] |
| std.err | 0.2954 | 0.2802 | 0.0140 |
| t-stat | 5.96 | 7.97 | 84.82 |
| p-value | 0 | 0 | 0 |

**Table S9: Relationship between Temperature, Relative Humidity, and *R* Value: Robustness Check with a social distancing dummy variable for the U.S. Counties.**

U.S. states lifted stay-at-home orders, namely a series of social distancing policies, at different times. This table shows the regression results for the U.S. Counties with an additional dummy explanatory variable recording whether the state where a county is located already lifted a stay-at-home order. The regression is estimated by the Fama-MacBeth approach.

|  | Overall | Before Lockdown (April 7) | After Lockdown (April 7) |
|---|---|---|---|
| $R^2$ | 0.1201 | 0.1403 | 0.0956 |
| **Temperature** | | | |
| coef | -0.0158 | -0.01988 | -.01092 |
| 95%CI | [-0.0246,-0.0071] | [-0.0300,-0.0097] | [-0.0265,0.0047] |
| std.err | 0.0043 | 0.0049 | 0.0074 |
| t-stat | -3.65 | -4.07 | -1.47 |
| p-value | 0.0007 | 0.0005 | 0.159 |
| **Relative Humidity** | | | |
| coef | -0.0050 | -0.0080 | -0.0014 |
| 95%CI | [-0.0104,0.0004] | [-0.0151,-0.0010] | [-0.0026,0.0002] |
| std.err | 0.0027 | 0.0034 | 0.0006 |
| t-stat | -1.88 | -2.37 | -2.46 |
| p-value | 0.067 | 0.027 | 0.024 |
| **Population Density** | | | |
| coef | 4.56e-06 | 7.77e-06 | 6.89e-07 |
| 95%CI | [-1e-5,2e-2] | [-2.53e-5,4.08e-5] | [-1.10e-6,2.48e-6] |
| std.err | 8.34e-06 | 1.59e-05 | 8.53e-07 |
| t-stat | 0.55 | 0.49 | 0.81 |
| p-value | 0.587 | 0.631 | 0.430 |
| **Percentage over 65** | | | |
| coef | -0.948 | -1.1645 | -0.6851 |
| 95%CI | [-1.3747,-0.5205] | [-1.8362,-0.4927] | [-1.0610,-0.3092] |

|  | Overall | Before Lockdown (April 7) | After Lockdown (April 7) |
|---|---|---|---|
| std.err | 0.2115 | 0.3239 | 0.1789 |
| t-stat | -4.48 | -3.60 | -3.83 |
| p-value | 6e-5 | 0.002 | 0.001 |
| **Gini** | | | |
| coef | -1.8813 | -1.9719 | -1.7717 |
| 95%CI | [-3.5537,-0.2090] | [-4.5293,0.5855] | [-2.5073,-1.0360] |
| std.err | 0.8281 | 1.2331 | 0.3502 |
| t-stat | -2.27 | -1.60 | -5.06 |
| p-value | 0.028 | 0.124 | 8e-5 |
| **Socio-economic factor** | | | |
| coef | 0.0891 | 0.1321 | 0.0371 |
| 95%CI | [0.0372,0.1411] | [0.0835,0.1807] | [-0.0048,0.0790] |
| std.err | 0.0257 | 0.02343 | 0.0200 |
| t-stat | 3.47 | 5.64 | 1.86 |
| p-value | 0.001 | 1e-05 | 0.079 |
| **No. of ICU beds per capita** | | | |
| coef | -0.0096 | -0.0084 | -0.0111 |
| 95%CI | [-0.0235,0.0043] | [-0.0301,0.0133] | [-0.0172,-0.0050] |
| std.err | 0.0069 | 0.0104 | 0.0029 |
| t-stat | -1.40 | -0.80 | -3.83 |
| p-value | 0.169 | 0.430 | 0.001 |
| **Fraction of maximum moving distance over normal time** | | | |
| coef | 0.0041 | 0.0031 | 0.0054 |
| 95%CI | [0.0016,0.0066] | [-0.0004,0.0067] | [0.0043,0.0065] |
| std.err | 0.0012 | 0.0017 | 0.0005 |
| t-stat | 3.35 | 1.82 | 10.25 |
| p-value | 0.002 | 0.082 | 0 |
| **Home-stay minutes** | | | |
| coef | 0.0003 | 0.0007 | -0.0002 |

|                    | Overall              | Before Lockdown (April 7) | After Lockdown (April 7) |
|--------------------|----------------------|---------------------------|--------------------------|
| 95%CI              | [-0.0002,0.0007]     | [0.0004,0.0010]           | [-0.0004,-3e-05]         |
| std.err            | 0.0002               | 0.0002                    | 9e-5                     |
| t-stat             | 1.33                 | 4.73                      | -2.42                    |
| p-value            | 0.191                | 0.0001                    | 0.026                    |
| **Latitude**       |                      |                           |                          |
| coef               | -0.0182              | -0.0348                   | 0.0018                   |
| 95%CI              | [-0.0371,0.0007]     | [-0.0510,-0.0185]         | [-0.0188,0.0225]         |
| std.err            | 0.0094               | 0.0078                    | 0.0098                   |
| t-stat             | -1.95                | -4.43                     | 0.19                     |
| p-value            | 0.058                | 0.0002                    | 0.854                    |
| **Longitude**      |                      |                           |                          |
| coef               | 0.0069               | 0.0103                    | 0.0029                   |
| 95%CI              | [0.0033,0.0106]      | [0.0082,0.0124]           | [0.0008,0.0050]          |
| std.err            | 0.0018               | 0.0010                    | 0.0010                   |
| t-stat             | 3.82                 | 10.13                     | 2.85                     |
| p-value            | 0.0005               | 0                         | 0.011                    |
| **Stay-at-home order** |                  |                           |                          |
| coef               | 0.0199               | 0.0939                    | -0.0695                  |
| 95%CI              | [-0.0651,0.1049]     | [0.0199,0.1678]           | [-0.13026,-0.088]        |
| std.err            | 0.0421               | 0.0356                    | 0.0289                   |
| t-stat             | 0.47                 | 2.63                      | -2.40                    |
| p-value            | 0.638                | 0.015                     | 0.027                    |
| **const**          |                      |                           |                          |
| coef               | 1.7395               | 2.1976                    | 1.1850                   |
| 95%CI              | [1.1800,2.2989]      | [1.6645,2.7306]           | [1.1695,1.2005]          |
| std.err            | 0.2770               | 0.2570                    | 0.0074                   |
| t-stat             | 6.28                 | 8.55                      | 160.27                   |
| p-value            | 0                    | 0                         | 0                        |

**Table S10: Relationship between Temperature, Relative Humidity, and *R* Value: Robustness Check with spatial random effect of Chinese cities.**

Spatial random effects are introduced in first step of Fama-Macbeth regression to account for spatial correlation. The neighborhood structure is calculated from the Earth distances between cities.

|  | Overall | Before Lockdown (Jan 24) | After Lockdown (Jan 24) |
|---|---|---|---|
| **Temperature** | | | |
| coef | -0.0212 | -0.0269 | -0.0196 |
| 95%CI | [-0.0361, -0.0063] | [-0.0429, -0.0108] | [-0.0377, -0.0016] |
| std.err | 0.0072 | 0.0058 | 0.0085 |
| t-stat | -2.96 | -4.65 | -2.30 |
| p-value | 0.007 | 0.010 | 0.034 |
| **Relative Humidity** | | | |
| coef | -0.0045 | -0.0074 | -0.0037 |
| 95%CI | [-0.0090, -0.00003] | [-0.0103, -0.0044] | [-0.0091, 0.0017] |
| std.err | 0.0022 | 0.0011 | 0.0026 |
| t-stat | -2.09 | -6.90 | -1.46 |
| p-value | 0.049 | 0.002 | 0.162 |
| **Population Density** | | | |
| coef | 0.0257 | 0.1059 | 0.0034 |
| 95%CI | [-0.0197, 0.0711] | [0.0208, 0.1911] | [-0.0200, 0.0268] |
| std.err | 0.0219 | 0.0307 | 0.0111 |
| t-stat | 1.17 | 3.45 | 0.31 |
| p-value | 0.253 | 0.026 | 0.764 |
| **Percentage over 65** | | | |
| coef | 0.0783 | 0.2110 | 0.0415 |
| 95%CI | [-1.5748, 1.7315] | [-1.1675, 1.5894] | [-2.0603, 2.1432] |
| std.err | 0.7971 | 0.4965 | 0.9962 |
| t-stat | 0.10 | 0.42 | 0.04 |

|  | Overall | Before Lockdown (Jan 24) | After Lockdown (Jan 24) |
|---|---|---|---|
| p-value | 0.923 | 0.693 | 0.967 |
| **GDP per capita** | | | |
| coef | -0.0022 | -0.0155 | 0.0015 |
| 95%CI | [-0.0203, 0.0159] | [-0.0262, -0.0048] | [-0.0187, 0.0218] |
| std.err | 0.0087 | 0.0038 | 0.0096 |
| t-stat | -0.25 | -4.04 | 0.16 |
| p-value | 0.805 | 0.016 | 0.876 |
| **No. of doctors** | | | |
| coef | -0.0056 | -0.0101 | -0.0044 |
| 95%CI | [-0.0083, -0.0030] | [-0.0163, -0.0039] | [-0.0059, -0.0029] |
| std.err | 0.0013 | 0.0022 | 0.0007 |
| t-stat | -4.40 | -4.52 | -6.10 |
| p-value | 0.0003 | 0.011 | 0.0002 |
| **Drop of BMI** | | | |
| coef | 0.2327 | -0.3903 | 0.4057 |
| 95%CI | [-0.3638, 0.8291] | [-0.6699, -0.1106] | [-0.2111, 1.0225] |
| std.err | 0.2876 | 0.1007 | 0.2924 |
| t-stat | 0.81 | -3.87 | 1.39 |
| p-value | 0.427 | 0.018 | 0.183 |
| **Inflow population from Wuhan** | | | |
| coef | -0.0028 | -0.0001 | -0.0035 |
| 95%CI | [-0.0055, -0.00004] | [-0.0011, 0.0008] | [-0.0063, -0.0007] |
| std.err | 0.0013 | 0.0003 | 0.0013 |
| t-stat | -2.11 | -0.43 | -2.62 |
| p-value | 0.047 | 0.688 | 0.018 |
| **Latitude** | | | |
| coef | 0.0063 | 0.0076 | 0.0059 |
| 95%CI | [-0.0161, 0.0286] | [-0.0191, 0.0343] | [-0.0221, 0.0339] |
| std.err | 0.0108 | 0.0096 | 0.0133 |

|  | Overall | Before Lockdown (Jan 24) | After Lockdown (Jan 24) |
|---|---|---|---|
| t-stat | 0.58 | 0.79 | 0.44 |
| p-value | 0.566 | 0.472 | 0.662 |
| **Longitude** | | | |
| coef | -0.0100 | -0.0258 | -0.0056 |
| 95%CI | [-0.0195, -0.0006] | [-0.0514, -0.0003] | [-0.0141, 0.0028] |
| std.err | 0.0046 | 0.0092 | 0.0040 |
| t-stat | -2.20 | -2.81 | -1.40 |
| p-value | 0.039 | 0.048 | 0.178 |
| **const** | | | |
| coef | 1.1002 | 2.1148 | 0.8183 |
| 95%CI | [0.5229, 1.6774] | [1.5587, 2.6710] | [0.5551, 1.0815] |
| std.err | 0.2784 | 0.2003 | 0.1247 |
| t-stat | 3.95 | 10.56 | 6.56 |
| p-value | 0.001 | 0 | 0.0002 |

**Table S11: Relationship between Temperature, Relative Humidity, and *R* Value: Robustness Check with spatial random effect of the U.S. counties.**

Spatial random effects are introduced in first step of Fama-Macbeth regression to account for spatial correlation. The neighborhood structure is calculated from the Earth distances between counties.

|  | Overall | Before Lockdown (April 7) | After Lockdown (April 7) |
|---|---|---|---|
| **Temperature** | | | |
| coef | -0.0136 | -0.0135 | -0.0136 |
| 95%CI | [-0.0215, -0.0057] | [-0.0236, -0.0034] | [-0.0280, 0.0007] |
| std.err | 0.0039 | 0.0049 | 0.0068 |
| t-stat | -3.46 | -2.78 | -2.00 |
| p-value | 0.001 | 0.011 | 0.061 |
| **Relative Humidity** | | | |
| coef | -0.0052 | -0.0072 | -0.0029 |
| 95%CI | [-0.0095, -0.0010] | [-0.0130, -0.0014] | [-0.0042, -0.0016] |
| std.err | 0.0021 | 0.0028 | 0.0006 |
| t-stat | -2.51 | -2.57 | -4.59 |
| p-value | 0.016 | 0.017 | 0.0003 |
| **Population Density** | | | |
| coef | 3.26e-8 | 2.98e-6 | -3.54e-6 |
| 95%CI | [-0.00002, 0.00002] | [-0.00003, 0.00004] | [-5.13e-6, -1.95e-6] |
| std.err | 8.58e-6 | 0.00002 | 7.57e-7 |
| t-stat | 0.00 | 0.18 | -4.67 |
| p-value | 0.997 | 0.858 | 0.0002 |
| **Percentage over 65** | | | |
| coef | -0.7988 | -1.0894 | -0.4471 |
| 95%CI | [-1.4330, -0.1647] | [-2.0771, -0.1017] | [-0.7620, -0.1322] |
| std.err | 0.3140 | 0.4763 | 0.1499 |
| t-stat | -2.54 | -2.29 | -2.98 |

|  | Overall | Before Lockdown (April 7) | After Lockdown (April 7) |
|---|---|---|---|
| p-value | 0.015 | 0.032 | 0.008 |
| **Gini** | | | |
| coef | -1.8186 | -2.2916 | -1.2460 |
| 95%CI | [-3.3837, -0.2534] | [-4.5288, -0.0543] | [-2.1425, -0.3495] |
| std.err | 0.7750 | 1.0788 | 0.4267 |
| t-stat | -2.35 | -2.12 | -2.92 |
| p-value | 0.024 | 0.045 | 0.009 |
| **Socio-economic factor** | | | |
| coef | 0.1131 | 0.1480 | 0.0708 |
| 95%CI | [0.0682, 0.1580] | [0.0903, 0.2056] | [0.0451, 0.0965] |
| std.err | 0.0222 | 0.0278 | 0.0122 |
| t-stat | 5.08 | 5.32 | 5.78 |
| p-value | 0.0002 | 0.0002 | 0.0002 |
| **No. of ICU beds per capita** | | | |
| coef | -0.0092 | -0.0127 | -0.0050 |
| 95%CI | [-0.0238, 0.0054] | [-0.0359, 0.0105] | [-0.0101, 0.0002] |
| std.err | 0.0072 | 0.0112 | 0.0025 |
| t-stat | -1.27 | -1.14 | -2.01 |
| p-value | 0.210 | 0.267 | 0.059 |
| **Fraction of maximum moving distance over normal time** | | | |
| coef | 0.0040 | 0.0024 | 0.0059 |
| 95%CI | [0.0012, 0.0068] | [-0.0014, 0.0063] | [0.0054, 0.0064] |
| std.err | 0.0014 | 0.0019 | 0.0002 |
| t-stat | 2.93 | 1.30 | 25.03 |
| p-value | 0.005 | 0.207 | 0 |
| **Home-stay minutes** | | | |
| coef | 0.0003 | 0.0005 | 0.00002 |
| 95%CI | [0.00002, 0.0006] | [0.0001, 0.0009] | [-0.0002, 0.0002] |
| std.err | 0.0001 | 0.0002 | 0.0001 |

|                | **Overall**        | **Before Lockdown (April 7)** | **After Lockdown (April 7)** |
| -------------- | ------------------ | ----------------------------- | ---------------------------- |
| t-stat         | 2.15               | 2.81                          | 0.19                         |
| p-value        | 0.038              | 0.010                         | 0.851                        |
| **Latitude**   |                    |                               |                              |
| coef           | -0.0152            | -0.0278                       | -0.00004                     |
| 95%CI          | [-0.0308, 0.0003]  | [-0.0423, -0.0133]            | [-0.0208, 0.0207]            |
| std.err        | 0.0077             | 0.0070                        | 0.0099                       |
| t-stat         | -1.98              | -3.97                         | -0.00                        |
| p-value        | 0.055              | 0.001                         | 0.997                        |
| **Longitude**  |                    |                               |                              |
| coef           | 0.0060             | 0.0084                        | 0.0032                       |
| 95%CI          | [0.0033, 0.0088]   | [0.0064, 0.0104]              | [0.0015, 0.0049]             |
| std.err        | 0.0014             | 0.0010                        | 0.0008                       |
| t-stat         | 4.45               | 8.78                          | 3.86                         |
| p-value        | 0.0003             | 0                             | 0.001                        |
| **const**      |                    |                               |                              |
| coef           | 1.7377             | 2.2018                        | 1.1759                       |
| 95%CI          | [1.1715, 2.3039]   | [1.6623, 2.7413]              | [1.1594, 1.1923]             |
| std.err        | 0.2803             | 0.2601                        | 0.0078                       |
| t-stat         | 6.20               | 8.46                          | 150.10                       |
| p-value        | 0                  | 0                             | 0                            |